\documentclass[a4paper,11pt]{article}
\pdfoutput=1 

\usepackage{jcappub} 

\usepackage[T1]{fontenc} 
\usepackage[dvipsnames]{xcolor}
\usepackage{tikz}
\usepackage{subcaption}
\usepackage{graphicx}
\usepackage[abs]{overpic}
\usepackage{soul}

\def\bi#1{\hbox{\boldmath{$#1$}}}

\usepackage{amssymb, amsmath, epsfig, natbib}

\def\apjl{ApJL}
\def\apjs{ApJS}

\def\aap{{A\&A}}

\author[a,b]{B. Horowitz}
\author[a,b]{U. Seljak}
\author[a]{G. Aslanyan}

\affiliation[a]{Berkeley Center for Cosmological Physics and Department of Physics, University of California,
Berkeley,  CA 94720}
\affiliation[b]{Lawrence Berkeley National Laboratory, Berkeley, CA 94720}

\emailAdd{bhorowitz@berkeley.edu, useljak@berkeley.edu}

\title{Efficient Optimal Reconstruction of Linear Fields and Band-powers from Cosmological Data}


\abstract{
We present an efficient implementation of Wiener filtering of real-space linear field and optimal quadratic estimator of its power spectrum Band-powers. We first recast the field reconstruction into an optimization problem, which we solve using quasi-Newton optimization. 
We then recast the power spectrum estimation into the field marginalization problem, from which we obtain an 
expression that depends on the field reconstruction solution and a determinant term. We develop a novel simulation based 
method for the latter. We extend the simulations formalism to provide the covariance matrix for the power spectrum. 
We develop a flexible framework that can be used on a variety of cosmological fields and present results for a variety of test cases, using simulated examples of projected density fields, projected shear maps from galaxy lensing, and observed Cosmic Microwave Background (CMB) temperature anisotropies, with a wide range of map incompleteness and variable noise. For smaller cases where direct numerical inversion is possible, we show that our solution matches that created by direct Wiener Filtering at a fraction of the overall computation cost. 
Even more significant reduction of computational is achieved by this implementation of optimal quadratic 
estimator due to the fast evaluation of the Hessian matrix. 
This technique allows for accurate map and power spectrum reconstruction with complex masks and nontrivial noise properties.}

\begin{document}
\maketitle
\flushbottom

\section{Introduction}

A ubiquitous problem of modern astrophysics is the reconstruction of the underlying signal from observed, noisy, and incomplete data. For linear fields the Wiener filter \cite{wiener1949extrapolation,rybicki1992interpolation} is the gold standard for reconstructing the underlying signal, as it is ``optimal'' in the sense that it minimizes the variance. It has been used as the basis of cosmological analysis for both large scale structure \cite{fisher1995lss,Kitaura2008bayesian} and CMB \cite{Bunn1996cobe,Tegmark2003wmap}.

However, the Wiener filter requires one to take the inverse of the overall covariance matrix, which has a noise and a signal based components. Noise is typically diagonal in observed space, 
while signal is diagonal in harmonic (or Fourier) space. In general, the covariance matrix will not be diagonalizable in either basis and it will be computationally difficult to invert the matrix numerically for a realistic sized survey map. While it is possible to make simplifying assumptions, like homogeneous and isotropic noise (as in, for example \cite{komatsu2005}), it is possible to efficiently implement Wiener filter by using the well studied property that the Wiener filter is mathematically equivalent to maximum a posteriori (MAP), which in turn can be solved with fast linear algebra methods or optimization \cite{seljak1998cosmography}. 

This optimization can be performed using a variety of numerical techniques. Recent work used a messenger \cite{ElsnerWandelt2013, 2016Alsing} (or dual messenger \cite{KodiLavauxEtAl2017}) field which can be diagonalized in either basis in which to run an approximation scheme. These methods have been argued to perform well versus other approximation schemes, but there are situations 
where the messenger field is zero (such as zero noise field with mask) and the method fails. In addition, it has been argued that a suitably chosen preconditioned conjugate gradient technique might be faster in some instances \cite{dmfvspcg}. Preconditioned conjugate gradient techniques have performed well in the context of CMB map reconstruction \cite{2004Jewell,2017Seljebotn}, but require careful selection of the preconditioning scheme to achieve fast convergence. We therefore want to explore implementations that are both general and computationally efficient. 

For cosmological analysis, it is not only the field that is of interest but also the band-powers (such as power spectrum amplitudes) and their covariance matrix. Work towards estimating these quantities jointly with the underlying field has been done in the cosmic shear context \cite{2016Alsing,2017Alsing}, but it required sampling from the joint probability distribution which is computationally involved. In this work we assume flat prior on band-powers and hence examine maximum likelihood estimation (MLE) method, 
after marginalization over the field, to quickly compute these quantities for observations with complex noise and mask properties as well.

In particular, we examine three cases of cosmological interest: linear density reconstruction, cosmic shear (E mode estimation alone as well as joint E/B), and primary CMB temperature anisotropy reconstruction. The outline of the paper is as follows: we discuss our MAP/MLE for field and power spectrum estimation in \S\ref{sec:background}. In \S\ref{sec:examples}, we use these tools in a number of contexts, and compare the optimization approach with a numerically exact case in Subsection \ref{subsec:wf}. In \S\ref{sec:conclusions} we discuss our results and possible extensions of the work to analyze real data.

In Appendix \ref{app:wf} we review the exact Weiner Filter approach with relies on inversion of the full covariance matrix. In Appendix \ref{app:CMB_rec} we discuss the convergence criteria in the case of primary CMB reconstruction. In Appendix \ref{app:EB} we discuss the joint estimation of E and B fields in the context of cosmic shear (although the technique transfers directly to CMB E and B polarization reconstruction as well).

\section{Background}
\label{sec:background}
Here we summarize the optimization technique and standardize the notation. For a more through description, see \cite{seljak1998cosmography,2009simon,2011InformationFieldTheory,seljak2017towards}. 
We measure quantities $d(\bi{r}_i)$ at select positions, such as a reconstructed projected density, forming a full data vector, $\bi{d}$.  In general, this data vector will depend on a combination of underlying information about the field ( ``field coefficients") we wish to estimate, $\bi{s}$, and that which is pure noise, $\bi{n}$, i.e.
\begin{equation}
\bi{d}=\bi{R}\bi{s} + \bi{n}
\label{eq:ddecomp},
\end{equation}
where the $\bi{R}$ is the response matrix expressing how our measurement changes with the underlying information. We express the underlying two point function information in terms of covariance matrices, $\bi{S}=\langle \bi{s}\bi{s}^\dagger \rangle$, and $\bi{N}=\langle \bi{n}\bi{n}^\dagger \rangle$, for the estimated signal and noise components, respectively. We assume that these are uncorrelated with each other, i.e. $ \langle \bi{n}\bi{s}^\dagger \rangle = 0$; changes to this assumption are tractable but would require a redefinition of our underlying likelihood function and complicate the analysis since the noise would carry signal information. The correlation matrix of the data is therefore,
\begin{eqnarray}
\langle \bi{d}\bi{d}^\dagger \rangle \equiv C &=& \langle (\bi{R}\bi{s}+ \bi{n})(\bi{R}\bi{s}+ \bi{n})^\dagger \rangle \nonumber \\
&=& \langle (\bi{R}\bi{s} (\bi{R}\bi{s})^\dagger + \bi{n} \bi{n}^\dagger + \text{Cross Terms} \rangle \nonumber \\
&=& \bi{R}\bi{S}\bi{R}^\dagger  + \bi{N}.
\end{eqnarray}
Our signal covariance, $\bi{S}$, takes a diagonal form in Fourier space. The noise covariance, $\bi{N}$, is often approximately diagonal in data space, as there usually are no noise correlations between elements of the data vector. We can convert the 
covariance matrix into Fourier basis where signal covariance is diagonal, but 
this will lead to very strong off-diagonal terms of the noise matrix, in the presence of masked sky regions and/or variable noise properties. Therefore, our overall data correlation matrix cannot be diagonalized easily in either basis. 

We can re-express our covariance in terms of underlying band-powers, $\bi{\Theta}$, labeling each modes of interest to estimate as $\{1,\cdots,l,\cdots, l_{max}\}$, and the rest as $\{ l_{max+1}, \cdots, \infty \}$, and then expressing our correlation matrix as 

\begin{equation}
\bi{C} = \sum_l \Theta_l \bi{Q}_l + \bi{N}.
\end{equation}

This new $\bi{Q}_l= \bi{\Pi}_l \bi{R} \bi{R}^\dagger \bi{\Pi}_l$ basis is the projection (with projection operators $\bi{\Pi}_l$) of the response matrix $\bi{R}$ for each mode. The
band-power can correspond to averaging over spherical harmonic basis 
coefficients at a constant $l$, adding up $2l+1$ 
$m$ modes, but we can also average over more than one $l$.  

\subsection{Bandpowers posterior}

We now want to find the most probable set of bandpowers for a given set of measurements $\bi{d}$, assuming flat prior on band-powers. We thus parametrize the power spectrum as a function of 
these bandpowers $\bi{S}(\bi{\Theta})$. 
If we assume our modes are Gaussian we can express the likelihood function in the familiar form, i.e.
\begin{equation}
L(\bi{d}|\bi{\Theta}) = (2 \pi)^{-N/2} \det(\bi{C})^{-1/2} \exp{\left(-\frac{1}{2}\bi{d}^\dagger \bi{C}^{-1} \bi{d}\right)}.
\label{eq_like}
\end{equation}
Associated with the likelihood function and a parameter set {$\bi{\hat{\Theta}}$} which maximizes it, is the Hessian matrix \cite{BondJaffeEtAl98},
\begin{equation}
F_{ll'}=-\frac{\partial^2 \text{ln} L}{\partial\Theta_l \partial \Theta_{l'}}  .
\end{equation}
The inverse of the Hessian matrix can be interpreted as a local estimate of the covariance matrix of the parameters, i.e. 
\begin{equation}
\bi{F}^{-1} = \langle \bi{\Theta \Theta^\dagger} \rangle -\langle \bi{\Theta}\rangle \langle \bi{\Theta}\rangle^\dagger.
\label{eq:var}
\end{equation}
We now have the bandpower posterior in the Gaussian form, given by 
the mean $\bi{\hat{\Theta}}$ and the covariance matrix $\bi{F}^{-1}$.
To obtain the solution for the mean it is easiest to 
use Newton's second order method, which gives a quadratic estimator of the form \cite{Tegmark97}
\begin{equation}
\Theta_l = \frac{1}{2}\sum_{l'}F^{-1}_{ll'}(\bi{d}^\dagger \bi{C}^{-1} \bi{Q}_{l'} \bi{C}^{-1}\bi{d} - b_{l'}),
\label{eq:ql}
\end{equation}
where $b_l$ is a noise bias term that can be found by computing the ensemble average of the first term assuming $\theta_l = 0$ for all modes probed (i.e. $l< l_{\rm max}$), 
\begin{equation}
b_l = \text{tr} \left[ \bi{N} + \sum_{l_{\rm max}+1}^{\infty} (\Theta_l \bi{Q}_l)\bi{C}^{-1}\bi{Q}_l\bi{C}^{-1}\right].
\end{equation}
This is an implicit equation since $\bi{C}$ depends 
on $\Theta_l$, and needs iterations, as discussed further below. We refer below the power spectrum inside $\bi{C}$ as 
$\bi{S}_{\rm fid}$.

\subsection{MAP Field Reconstruction}
\label{subsec:ml}
In practice this analytical calculation requires the inversion of a large matrix, $\bi{C}$, which does not necessarily have properties that make inversion efficient (i.e. block diagonal or sparse) and will in general require $O(n^3)$ time for an $n \times n$ matrix. In the case of reconstructing the underlying density field for astronomical large surveys with $n$ pixels, this would be prohibitively computationally expensive for the foreseeable future. Instead, we will approach this as an optimization problem \cite{seljak2017towards}. We will not use \ref{eq_like}, and instead of expressing the likelihood of data given bandpowers we will work in terms of latent variables, writing the joint
distribution of $\bi{s}$ and $\bi{d}$,
\begin{equation}
p(\bi{s},\bi{d}|\bi{S}) = (2 \pi)^{-(N+M)/2} det(\bi{SN})^{-1/2} \exp{\left(-\frac{1}{2}\bi{s}^\dagger \bi{S}^{-1} \bi{s} + (\bi{d}-\bi{Rs})^\dag \bi{N}^{-1}(\bi{d}-\bi{Rs}) \right)},
\label{eq_likeb}
\end{equation}
and note that the minimum variance solution for the modes can be found by minimizing the loss function $\chi^2$,
\begin{equation}
\chi^2 = -2\ln p(\bi{s},\bi{d}|\bi{S})+c=\bi{s}^\dagger \bi{S}^{-1} \bi{s} + (\bi{d}-\bi{Rs})^\dag \bi{N}^{-1}(\bi{d}-\bi{Rs}),
\label{eq_chi}
\end{equation}
with respect to $\bi{s}$. 
Taylor expanding around $\bi{s_m}$ to second order, we have
\begin{equation}
\chi^2 = \chi^2(\bi{s_m}) + 2 \bi{g}(\bi{s}-\bi{s_m})+ (\bi{s}-\bi{s_m})^\dagger\bi{D}(\bi{s}-\bi{s_m}),
\label{eq_chib}
\end{equation}
with gradient function in terms of the derivative of the response function, $\bi{R}$, given as
\begin{equation}
\bi{g}=\frac{1}{2}\frac{\partial \chi^2}{\partial \bi{s}}=\bi{S}^{-1}\bi{s_m}-\bi{R}^\dag\bi{N}^{-1}(\bi{d}-\bi{Rs_m}).
\label{eq_cost}
\end{equation}
For the linear problems studied in this work, this derivative can be calculated analytically, but in other more involved cases (such as nonlinear structure formation) might be computationally involved as it would require intensive back-propagation. The solution where $\bi{g}=0$, and therefore a local extremum is found, will be denoted $\hat{\bi{s}}$, 
and is the maximum a posteriori solution (MAP). For linear problems it is the best 
possible solution in the sense to minimizing the variance. 

The curvature matrix $\bi{D}$ 
has the form
\begin{equation}
\bi{D}=\frac{1}{2}\frac{\partial^2 \chi^2}{\partial \bi{s}\partial \bi{s}}=\bi{S}^{-1}+\bi{R}^\dag\bi{N}^{-1}\bi{R}.
\label{eq_curv}
\end{equation}
However, in this work we will not explicitly evaluate it, as it is too 
large. Instead, we will use low rank approximation as performed by
L-BFGS quasi-Newton optimization method. We will use L-BFGS as the 
optimization method in this paper. 

The starting point for the optimization algorithm  does not play a significant role for linear problems as the posterior surface is convex and the true global minimum can always be found. In practice, for the examples in this work, we found no noticeable effects of the starting point 
in terms of convergence properties, i.e. required number of iterations. 

\subsection{Minimum Variance Estimation of the Power Spectrum}
\label{subsec:ps_est}

The result of the above optimization procedure is $\bi{\hat{s}}$, and is useful for creating maps, but has more information than needed for cosmological analysis. If our goal is to determine a set of summary statistics/band-power measurement, $\bi{\Theta}$, such as a power spectrum bandpowers, we need to marginalize over the latent variables, the modes 
$\bi{s}$. To do so we need to define a projection matrix $\bi{\Pi}_l$ around a fiducial power-spectrum $\bi{S}_{\rm fid}$ with associated band-powers $\Theta_{\rm fid}$, defined as
\begin{equation}
\left[\frac{\partial \bi{S}}{\partial \Theta_l}\right]_{\bi{S}_{\rm fid}}=\bi{\Pi}_l.
\label{pi}
\end{equation}
This fiducial power spectrum is a regularized version of the measured power spectrum, and is 
thus iterated upon: we start with some fiducial prior, which we then update if the data require us to do so. 
This process is regularized, i.e. we use a smooth version of the measured power spectrum, for example a 
power spectrum predicted by the cosmological parameters we are determining from these data. 

The true covariance can be written in terms of the projection operators:
\begin{equation}
\bi{S}=\sum_l \Theta_l\bi{\Pi}_l,
\end{equation}
where $\Delta \Theta_l$ is the difference of the band-powers to those of the fiducial model. For the cases studied in this work, the dependence of $\bi{S}$ on $\bi{\Theta}$ is linear so we can take

\begin{equation}
\bi{\Pi}_l=\frac{\bi{S}_{\rm fid}}{\Theta_l}, 
\label{pilin}
\end{equation}
i.e. the projection matrix takes the power spectrum per bin, $\bi{\Theta}_l$, to the full power spectrum, $\bi{S}$. 
Note that the choice of the fiducial model is important in that if it is sufficiently far away from the true model the result could be biased, but iteratively recalculating $\bi{S}_{\rm fid}$ with the solved new band-powers $\bi{\Theta}$ will provide an asymptotically more accurate reconstruction. In the cases of interest in this work, a single iteration was sufficient to provide an accurate reconstruction. In practice for examples in this work, we used the true power-spectrum with each power re-scaled by a random value between 0.01 and 1.0; however we tested various other schemes which all provide accurate reconstructions as long as no band-power was set identically to zero.



We are assuming flat prior for the bandpowers, so 
to compute the posterior distribution of band-powers we can write their (marginalized over $\bi{s}$) likelihood function to maximize as a second order expansion around the fiducial model 
\begin{equation}
\ln L(\bi{\Theta}_{\rm fid}+\Delta \bi{\Theta})=\ln L(\bi{\Theta}_{\rm fid})+\sum_l \left[{\partial \ln L(\bi{\Theta}) \over \partial \Theta_l} \right]_{\bi{\Theta}_{\rm fid}}\Delta \Theta_l+ {1 \over 2} \sum_{ll'}\left[{\partial^2 \ln L(\bi{\Theta}) \over \partial \Theta_l \partial \Theta_{l'}}\right]_{\bi{\Theta}_{\rm fid}}\Delta \Theta_l\Delta \Theta_{l'};
\label{llik}
\end{equation}
where we assume a flat prior on the band-powers. 

We define
\begin{equation}
E_l(\bi{S}_{\rm fid},\hat{\bi{s}})={1 \over 2}\hat{\bi{s}}^{\dag}\bi{S}_{\rm fid}^{-1}\bi{\Pi}_l\bi{S}_{\rm fid}^{-1}\hat{\bi{s}} = \frac{1}{2}\sum_{k_l}{\hat{s}_{k_l}^2 \over \Theta_{\rm{fid},l} S_{{\rm fid},k_l}},
\label{el}
\end{equation}
where in the last expression we define the sum over $k_l$ as the sum over all modes which contribute to band-power, $\Theta_l$, and in the last equality we made use of the diagonal property of the projection operators and fiducial power spectrum. Putting this together we find that the derivative of the likelihood can be expressed as \cite{seljak2017towards}
\begin{equation}
{\partial \ln L(\bi{\Theta}) \over \partial \Theta_l}=E_l
-b_l ,
\label{like_deriv}
\end{equation}
where we defined 
\begin{equation}
b_l = {1 \over 2}{\rm tr}\left[{\partial \det \ln (\bi{SN}) \over \partial \Theta_l}\right]_{\bi{S}_{\rm fid}}.
\label{bl}
\end{equation}
For the linear cases studied in this work, this term is often called the noise bias term.
However, 
it is worth remembering that this term's origin is the derivative of the log determinant of the 
product of the Hessian and the signal covariance matrices in equation 
\ref{bl} (since noise covariance derivative 
is zero). 
To find MLE we need to find the zero of Eq. \ref{like_deriv}, which we solve using Newton's method. To do this we define the Hessian matrix,
\begin{equation}
F_{ll'}=-{\partial^2 \ln L(\bi{\Theta}) \over \partial \Theta_l \partial \Theta_{l'}} ,
\label{fish1}
\end{equation}
which for linear models defines the Gaussian posterior assuming sufficient modes have been averaged over so that by central limit theorem we can describe the 
posterior as a multi-variate gaussian. The peak of the likelihood function can be found by setting the derivative of equation \ref{llik} with respect to $\bi{\Delta \Theta}$ to zero,
which upon inserting equation \ref{like_deriv} yields
\begin{equation}
(\bi{F}\Delta \hat{\bi{\Theta}})_l=
E_l
-b_{l}.
\label{eq:wfmv}
\end{equation}

\subsection{Estimation of the Noise Bias and the Hessian}
While the noise bias, $b_l$, and the Hessian matrix, $\bi{F}$, from Equation \ref{eq:wfmv} could be calculated exactly, this will involve inversion of large matrices, which is what we are trying to avoid by deriving the MAP via optimization techniques. Instead, we will perform a simulation based analysis motivated by the underlying definition of each of these terms.

In general, the maximum likelihood field, $\hat{\bi{s}}$, attained with the procedure described in Sec \ref{subsec:ml}, will have bias due to the presence of noise: when the 
noise is high the minimum variance estimator drives $\bi{s}$ to zero. In the case of cosmological density fields which have red power spectra (less power on small scales compared
to white noise), this will result in washing out the small scale power. See the figures in Sec \ref{sec:examples} for explicit examples.

To correct for this bias we need to  understand how our reconstruction responds to the presence of noise. For this we perform a simulation analysis wherein we generate a data realization
generated from a fiducial power spectrum, 
inject the noise and mask, perform the optimization and see how the presence of noise affects the reconstruction. Let us call the new data and noise realization data $\bi{d}_{s+n}$, with associated maximum likelihood reconstruction $\hat{s}_{s+n}$. The gradient of
equation \ref{like_deriv} has to vanish if evaluated at the fiducial model. 
The noise bias in this case can be found directly as
\begin{equation}
b_l= E_l(\bi{\Theta}_{\rm fid},\hat{\bi{s}}_{s+n}). 
\label{blmv}
\end{equation}
This quantity should be averaged over many realizations, but for the linear signal-dominated cases studied in this work we found even one realization was sufficient for an accurate reconstruction.

To calculate the Hessian matrix, we evaluate the gradient of equation \ref{like_deriv}
at two different fiducial model values, and use finite differentiation
\cite{seljak2017towards},
\begin{equation}
{F}_{ll'}\Delta \Theta_{l'}=
E_l(\bi{\Theta}_{\rm fid}+\Delta \Theta_{l'})
-E_l(\bi{\Theta}_{\rm fid}).
\label{fisher}
\end{equation}
Its inverse is the covariance matrix for the band-powers. This is in contrast to directly using linear algebra techniques to calculate the Hessian matrix (see Equation \ref{eq:wf_exact_f}) which would be numerically intractable for a realistic survey size. Using Equation \ref{fisher} in the linear case, one can calculate it at the cost of additional optimization step. Since this is a linear problem Hessian matrix equals Fisher 
information matrix and thus gives the smallest attainable errors on the parameters (Cram\' er-Rao theorem).\cite{kay1993fundamentals}
 
\subsection{Procedure Summary}

\begin{enumerate}
\item Initialize a Gaussian random field (the true signal field) with some underlying power-spectrum.
\item Apply the response operator to this field, and additional noise and masking terms. The output of this is the input data vector, $\bi{d}$.
\item An estimate of the underlying signal field is created through optimization as described in Subsection \ref{subsec:ml} to yield $\bi{\hat{s}}$.
\item An initial estimate of the band-powers is generated by taking the power spectra of the reconstructed map and binning.
\item To this band-power estimate, we apply the noise bias correction (estimated using Eq \ref{blmv}) and Hessian matrix (estimated using Eq \ref{fisher}) to provide an optimal reconstructed value of the band-powers given in Eq \ref{eq:wfmv}.
\end{enumerate}

\section{Example Cases}
\label{sec:examples}

Here we implement the above scheme in a number of simulated cosmological contexts to demonstrate its versatility and efficiency. For these cases, we set our convergence criteria to be $\epsilon \equiv \delta \chi^2 = 10^{-1}$; i.e. the optimization ends when the difference of the absolute chi-square values between iterations is $10^{-1}$ (typical value of $\chi^2$ is of order $2\times 10^5$
for the dimensionality used here). We note that this is the largest which would be recommended to use to be suitably similar to the exact solution. In Appendix \ref{app:CMB_rec} we discuss the choice of this criteria in the context of CMB reconstruction, but we have found it to be sufficient for all the example cases.

\subsection{Projected Density Field}
\begin{figure}[t]

\begin{center}
\begin{overpic}[width=0.32\textwidth]{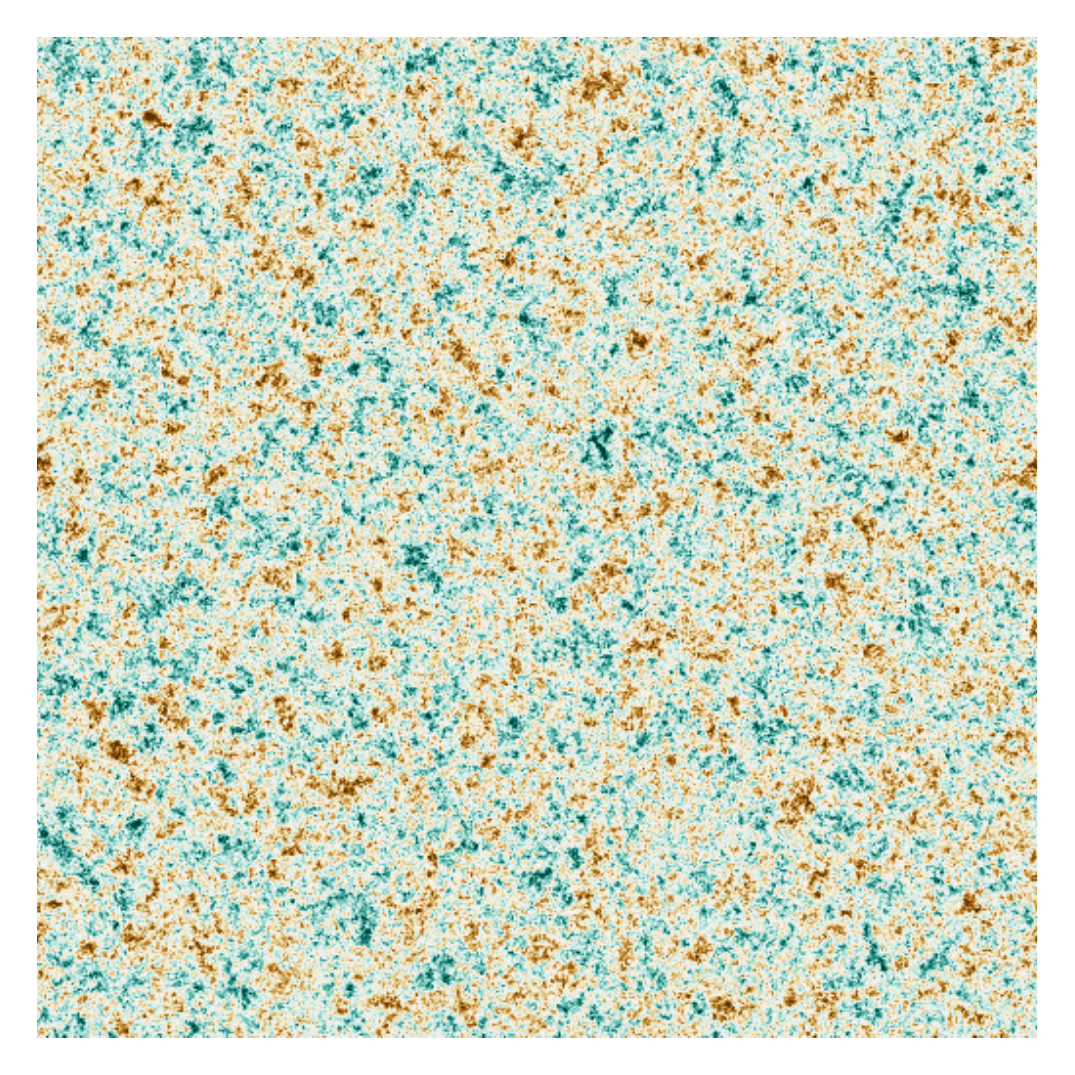}
\put(25,-10){\textsf{\scriptsize (a) Original Density Field }}
\end{overpic}
\begin{overpic}[width=0.32\textwidth]{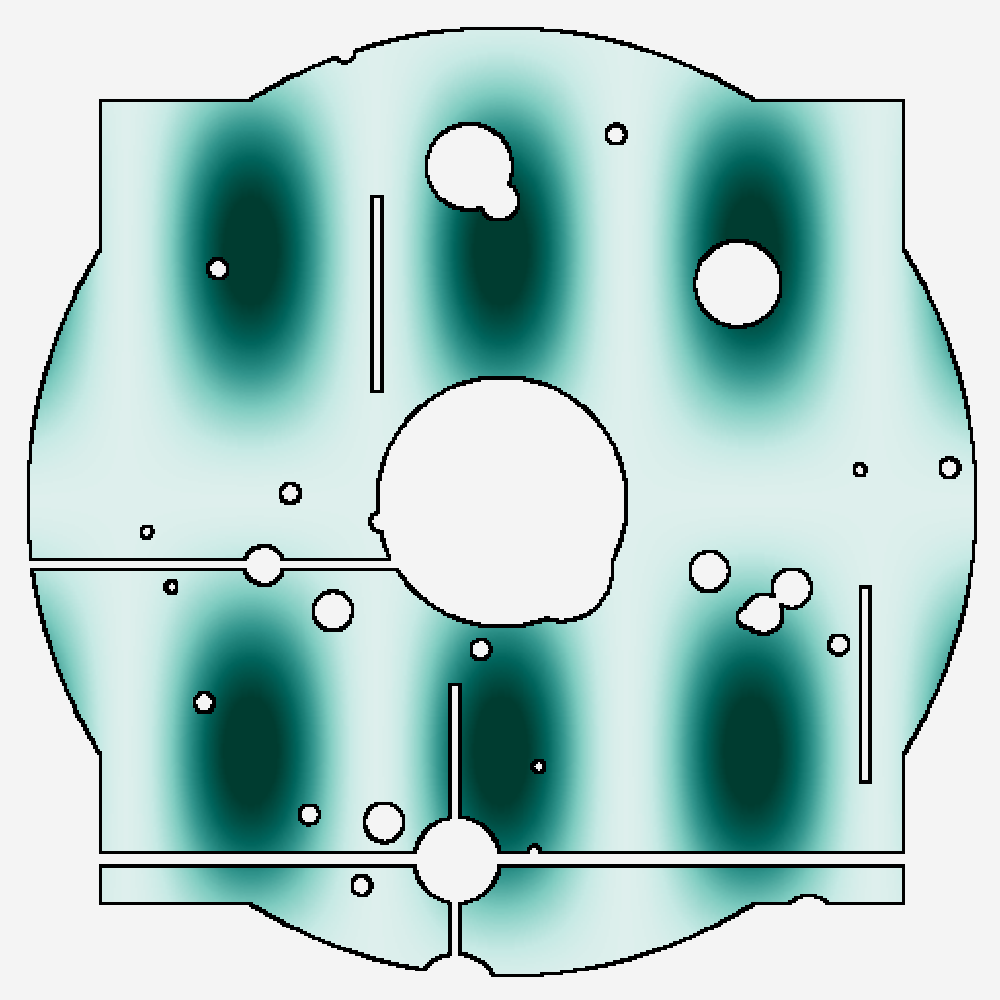}
\put(14,-10){\textsf{\scriptsize (b) Mask and Anisotropic Noise }}\end{overpic}
\vspace{1em}
\begin{overpic}[width=0.32\textwidth]{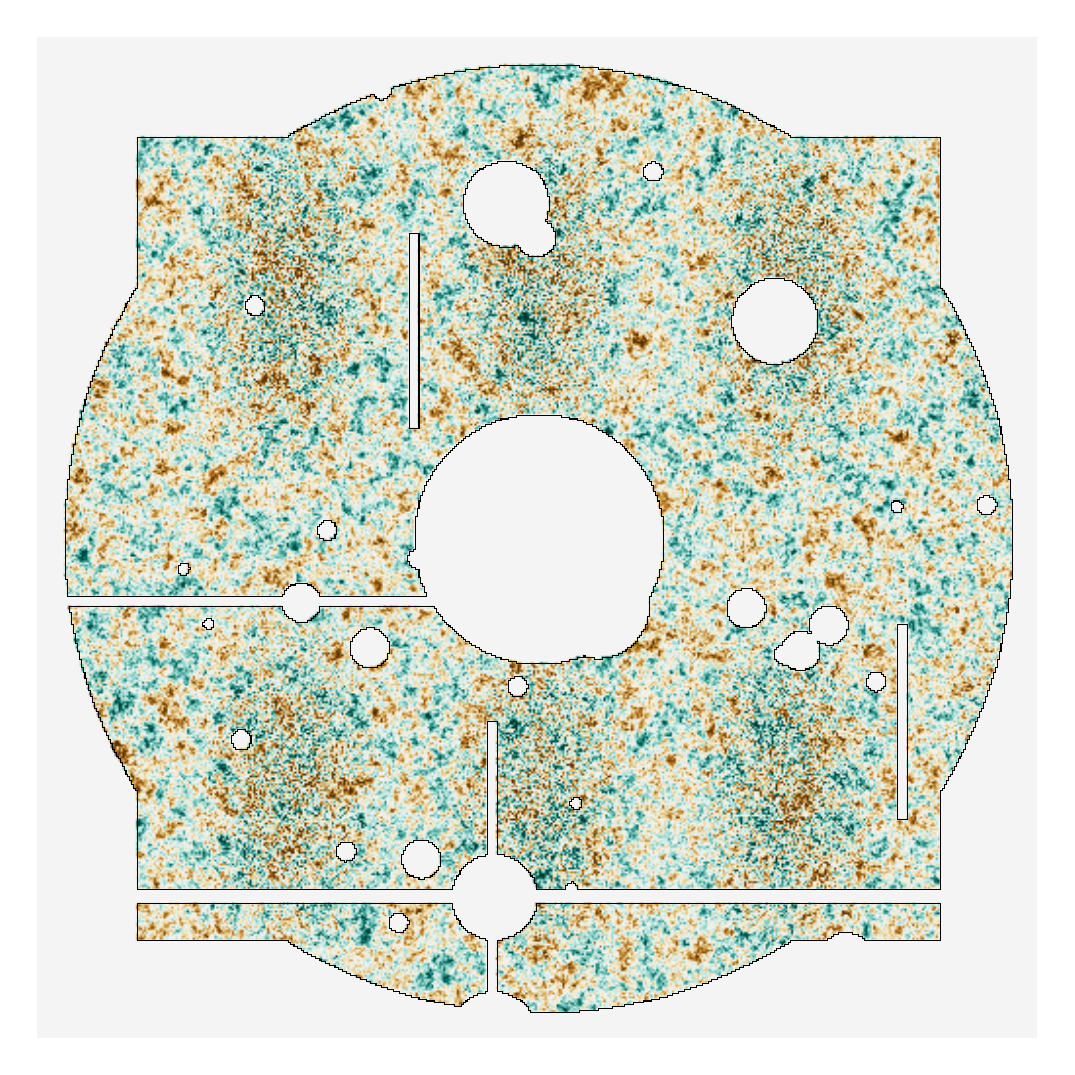}
\put(25,-10){\textsf{\scriptsize (c) Observed Density Field }}
\end{overpic}
\end{center}
\vspace{-0.4cm}

\begin{center}
\begin{overpic}[width=0.44\textwidth]{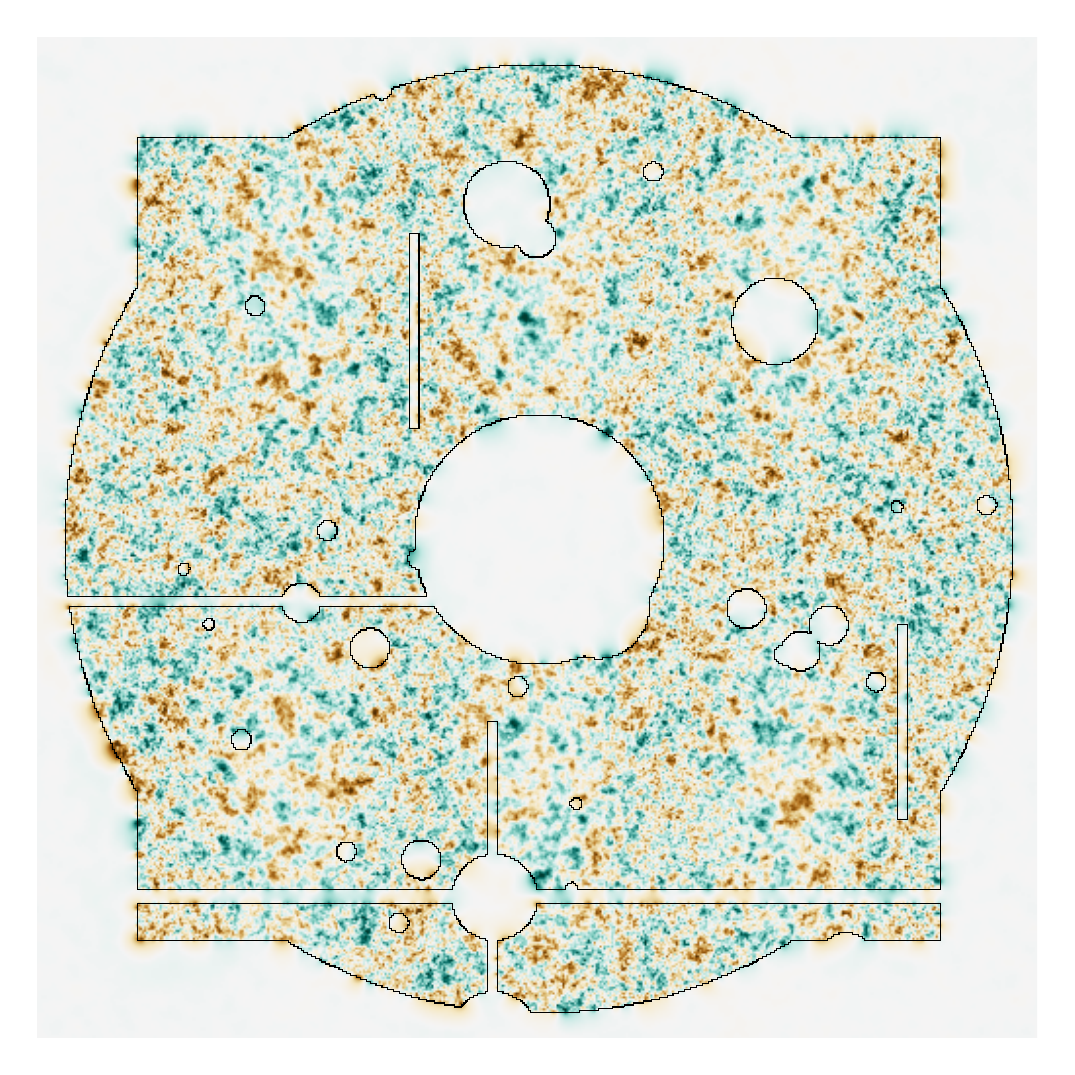}
\put(30,-10){\textsf{\scriptsize (d) Reconstructed Field }}
\end{overpic}
\vspace{1em}
\begin{overpic}[clip,trim={0.5cm 0.10cm 0 0cm},width=0.495\textwidth]{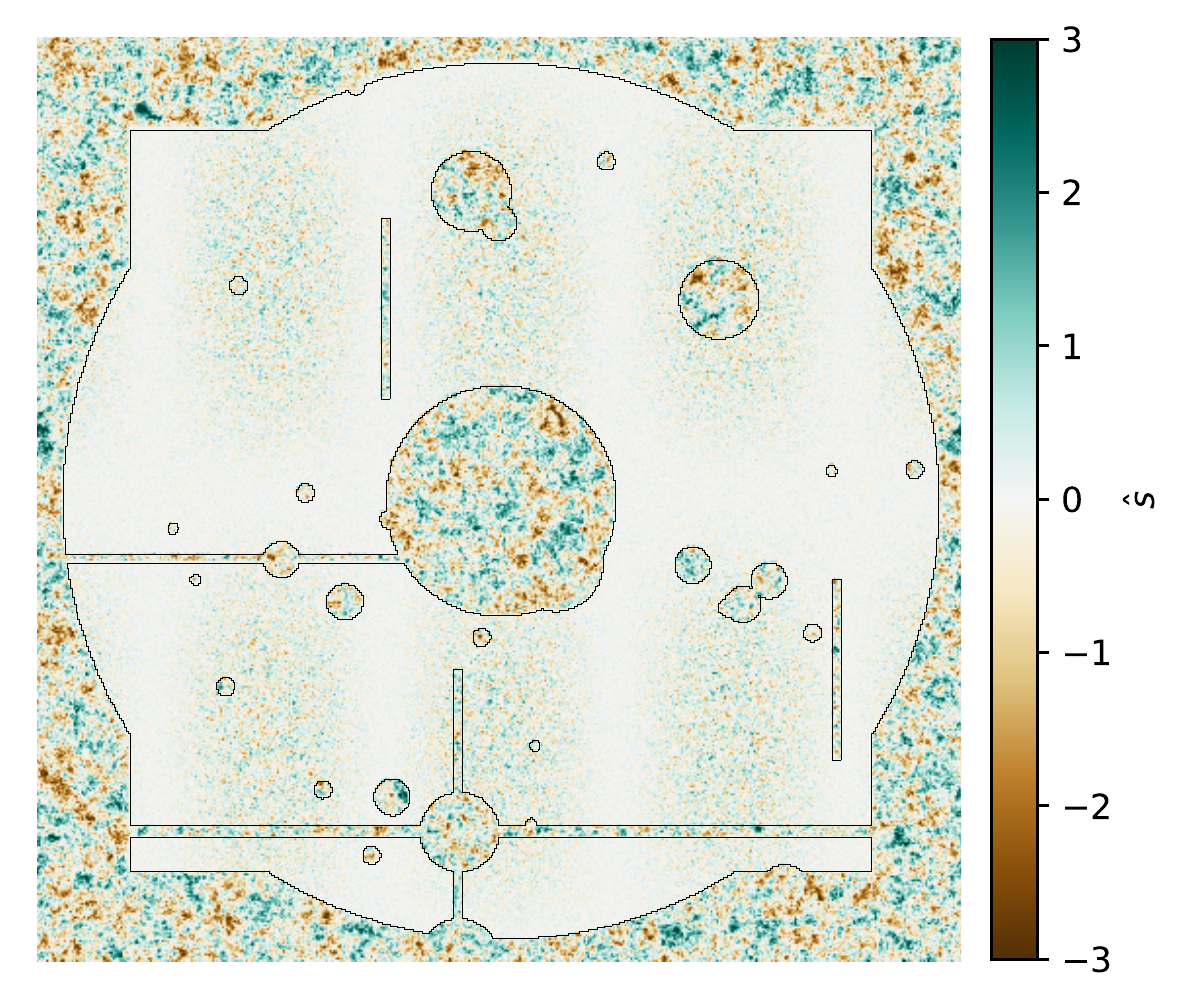}
\put(20,-10){\textsf{\scriptsize (e) True Field - Reconstructed }}
\end{overpic}
\vspace{1em}
\end{center}

\caption{\label{fig:512_d}
MAP density reconstruction for the 512x512 pixel case. Note that images (a), (c), (d),and (e) have same absolute color scale, while (b) shows the spatial variance of the noise properties. Color scale is normalized to show standard deviations away from mean.
}
\end{figure}

As our first case, we will look at reconstructing the density field from a noisy measurement of the density field. The starting measurement could come from  a variety of observations such as galaxy clustering or Lyman alpha forest tomography. For a complete analysis one would specify the response matrix $\bi{R}$ to go from the density to the observables which would include a model for the various biases present in the observations. In this case we assume the bias model is known and deal directly with the matter density field.

We generate a Gaussian random field with a power-spectrum given by
\begin{equation}
P(k) \propto \frac{k}{1+k^3}
\label{eq_simpleps}
\end{equation}
over a 2D, $L = 1380$ Mpc/h side-length box. This formula is chosen to provide a blue spectrum (significant small scale power) in order to compare later with the red Cosmic Microwave Background spectra. We introduce an anisotropic white noise over the field to simulate either irregularities in depth of a given survey or theoretical uncertainties in the underlying bias model. In Appendix \ref{app:wf} we demonstrate the validity of the L-BFGS optimization method in a small test case ($64 \times 64$ pixel) where it is also tractable to exactly invert the full covariance matrix numerically thereby providing validation of our maximum likelihood technique, while here we examine a more realistic $512 \times 512$ pixel map. We also use a realistic mask which includes foreground stars and other potential image defects.

Using the input power spectrum of Eq. \ref{eq_simpleps}, we generate a density field shown in Fig \ref{fig:512_d}(a), apply a mask and anisotropic noise shown in Fig \ref{fig:512_d}(b), which results in a mock observation in Fig \ref{fig:512_d}(c). We perform the minimization routine outlined in Sec. \ref{subsec:ml}, with the optimized map shown in Fig. \ref{fig:512_d}(d) with residuals shown in Fig \ref{fig:512_d}(e). Qualitatively the field is accurately reconstructed within the mask in the low-noise regions and is even able to reconstruct the larger scale modes right on the border within the masked region. However, as expected, the small scale modes within the high noise regions within the mask are poorly reconstructed since it is impossible to differentiate those modes in real space with the noise. In addition, small masked regions have very low residual error as there are sufficient, well sampled, nearby large scale modes to infer the regions value.
\begin{figure}
\begin{center}
\includegraphics[width = 0.7\textwidth]{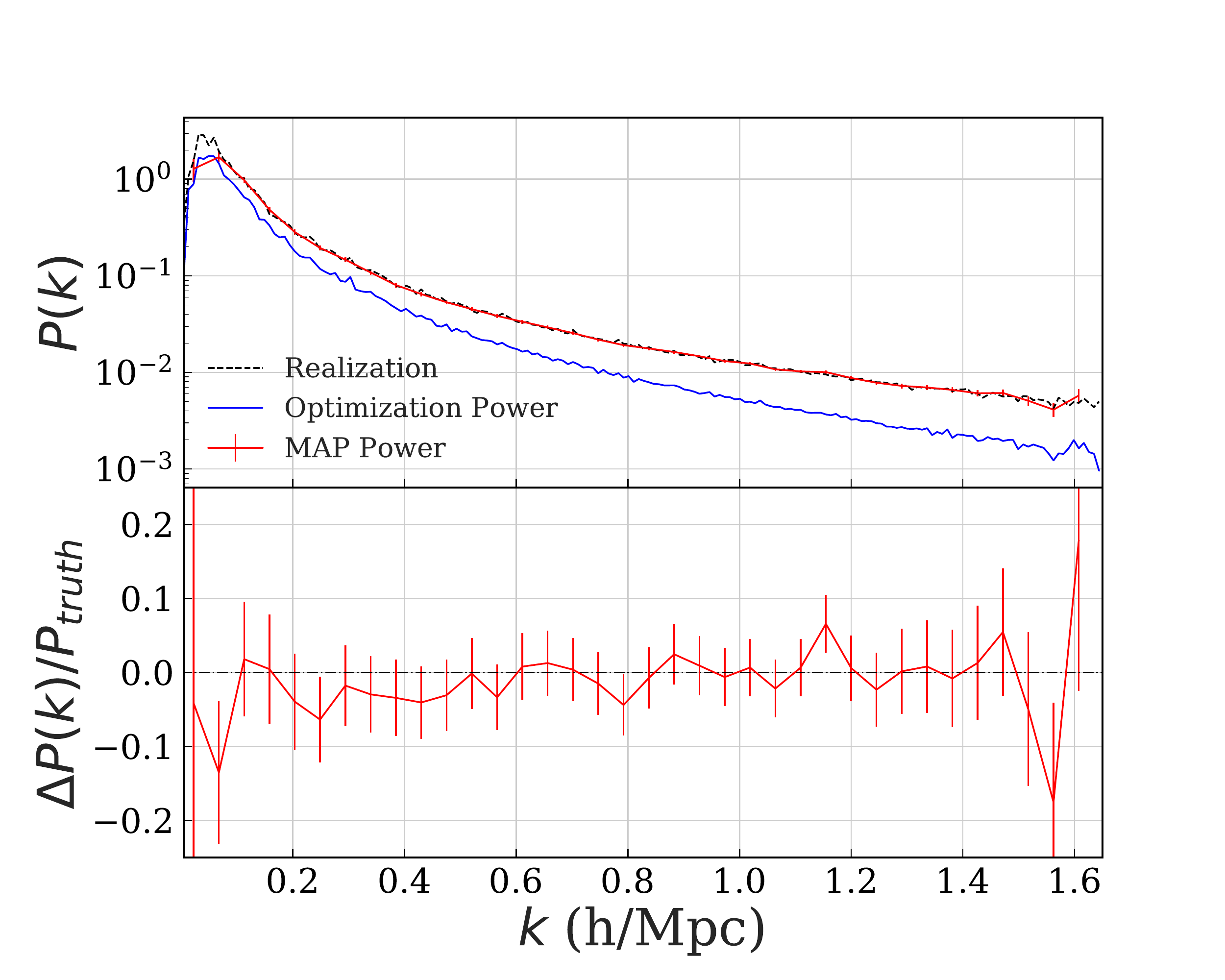}
\end{center}
\caption{\label{fig:pk512}Comparison of the maximum likelihood power spectrum attained from optimization versus the true power-spectrum of the region for the density field. Also shown is the importance of the noise bias correction (or, equivalently, the importance of the Hessian determinant). 
}
\end{figure}

In Fig \ref{fig:pk512} we show the comparison of the optimized result with the true power-spectra of the entire field. We also show the effect of the noise bias correction, which in this case is substantial as small scale power is washed out in the high noise regions as well as due to the masked regions. However, this power is recoverable using the analysis described in Section \ref{sec:background}.

\begin{figure}
\begin{center}
\includegraphics[width = 0.5\textwidth]{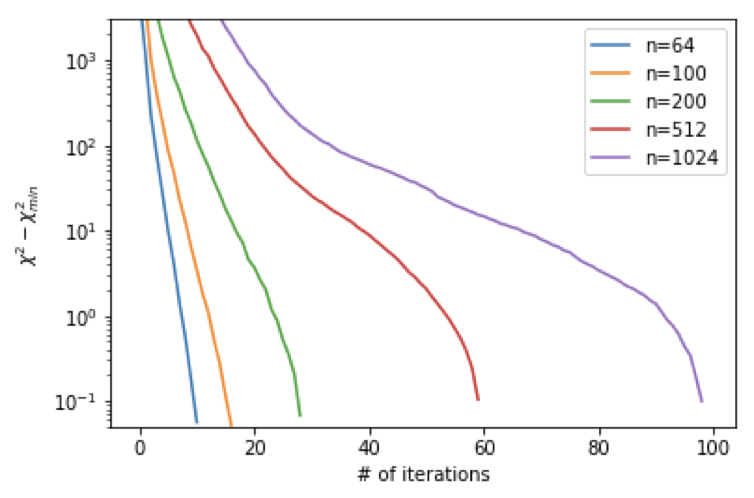}
\end{center}
\caption{\label{fig:density_iter_compare}
Convergence properties of the L-BFGS estimator as a function of the box-side dimension $n$.
}
\end{figure}

In Fig \ref{fig:density_iter_compare} we show how the number of iterations of the reconstruction algorithm scales with the box size, holding the noise per unit angle fixed. Each iteration requires a calculation of the loss function and the derivatives of the field which can be found analytically. Notice that the iteration number increases with the box size, but only a factor of a few when going from 
$64^2$ to $1024^2$. This does not include estimation of the noise bias and the Hessian matrix, which will depend on the number of bandpowers. As each row of the Hessian matrix requires an additional optimization, the true number of iteration will scale linearly with the number 
of bandpowers. In practice, since the Hessian matrix is very smooth and nearly translationally 
invariant (in this case, it is peaked on the diagonal and monotonically decreasing away from the diagonal), one simply needs to sample the matrix along a small number of rows and interpolate between them.

\subsection{Wiener Filtering vs. MAP Projected Density Example}
\label{subsec:wf}

Direct numerical evaluation of Wiener filtering is computationally expensive as it requires the direct inversion of a matrix with the square of the number of pixels in the survey (see App \ref{app:wf}), so we specialize our direct comparison to a small $64 \times 64$ pixel image.

\begin{figure}[t]
\begin{center}
\begin{overpic}[width=0.32\textwidth]{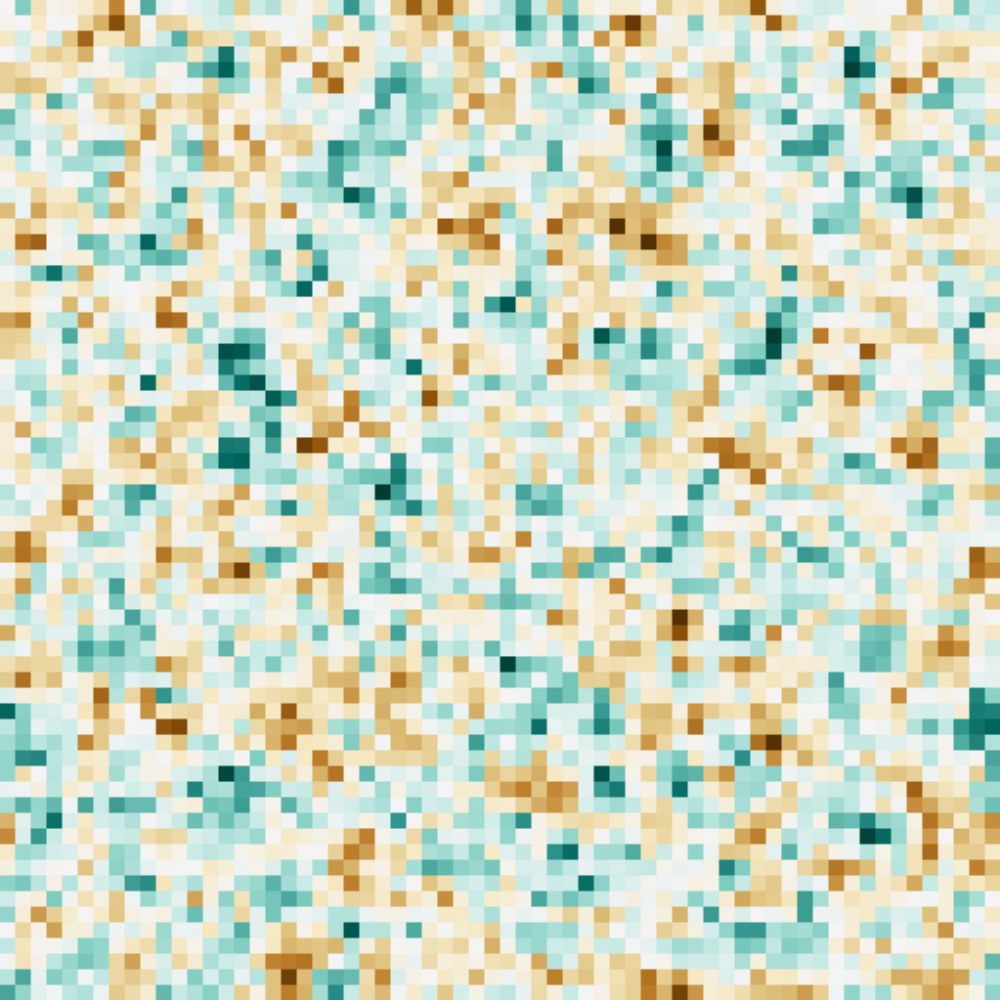}
\put(25,-10){\textsf{\scriptsize (a) Original Density Field }}
\end{overpic}
\begin{overpic}[width=0.32\textwidth]{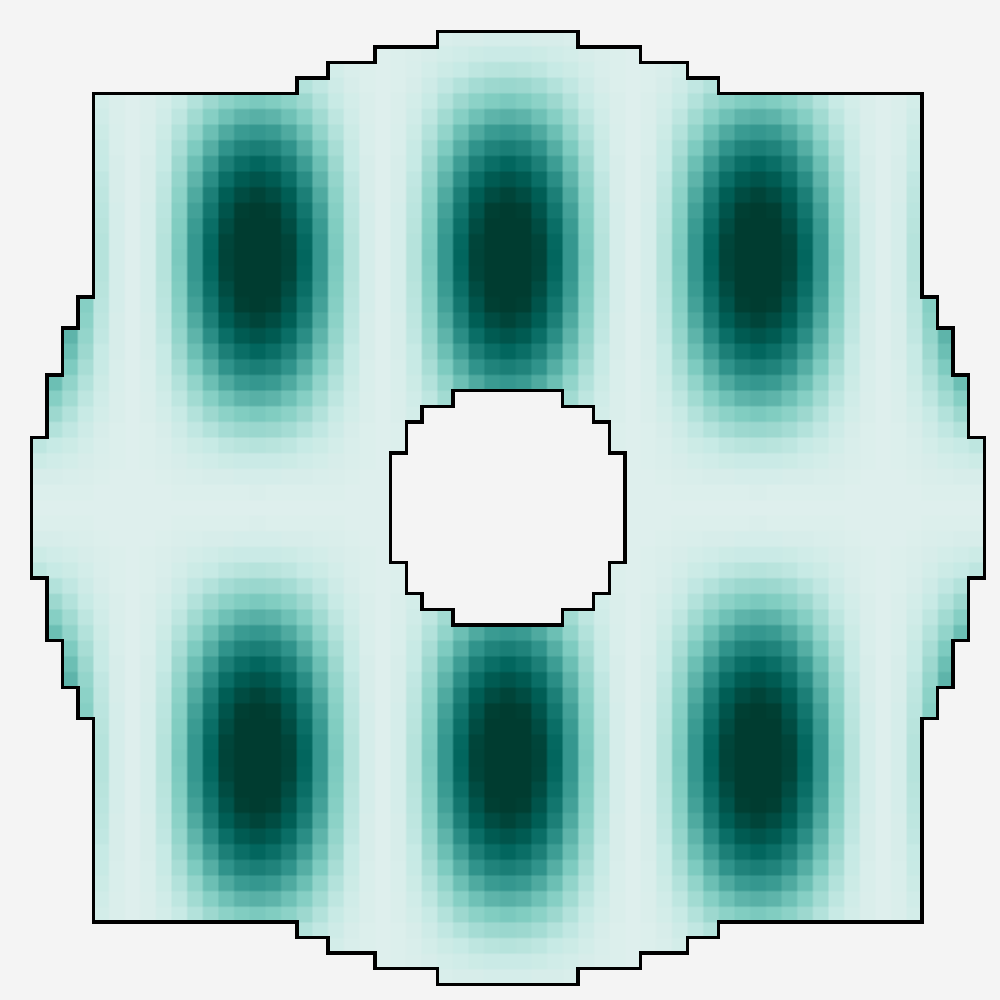}
\put(14,-10){\textsf{\scriptsize (b) Mask and Anisotropic Noise }}\end{overpic}
\vspace{1em}
\begin{overpic}[width=0.32\textwidth]{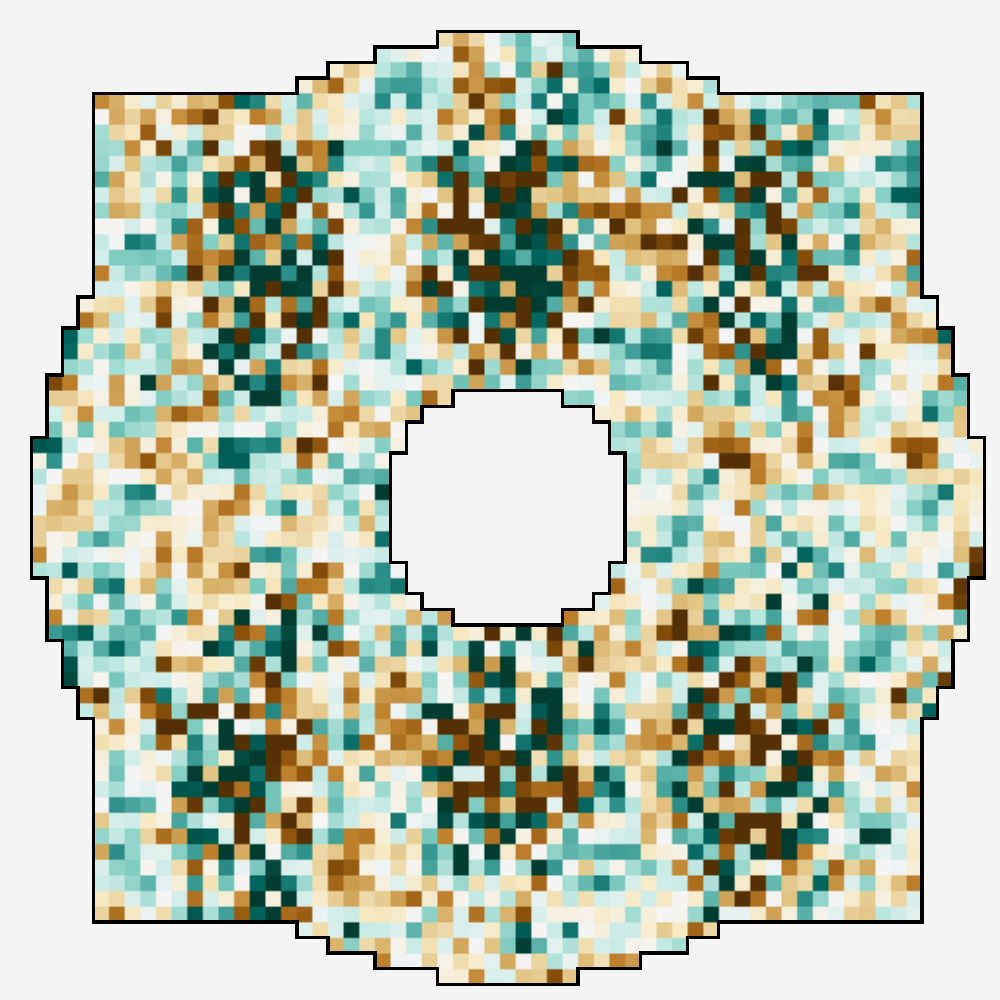}
\put(25,-10){\textsf{\scriptsize (c) Observed Density Field }}
\end{overpic}
\end{center}
\vspace{-0.4cm}

\begin{center}
\begin{overpic}[width=0.410\textwidth]{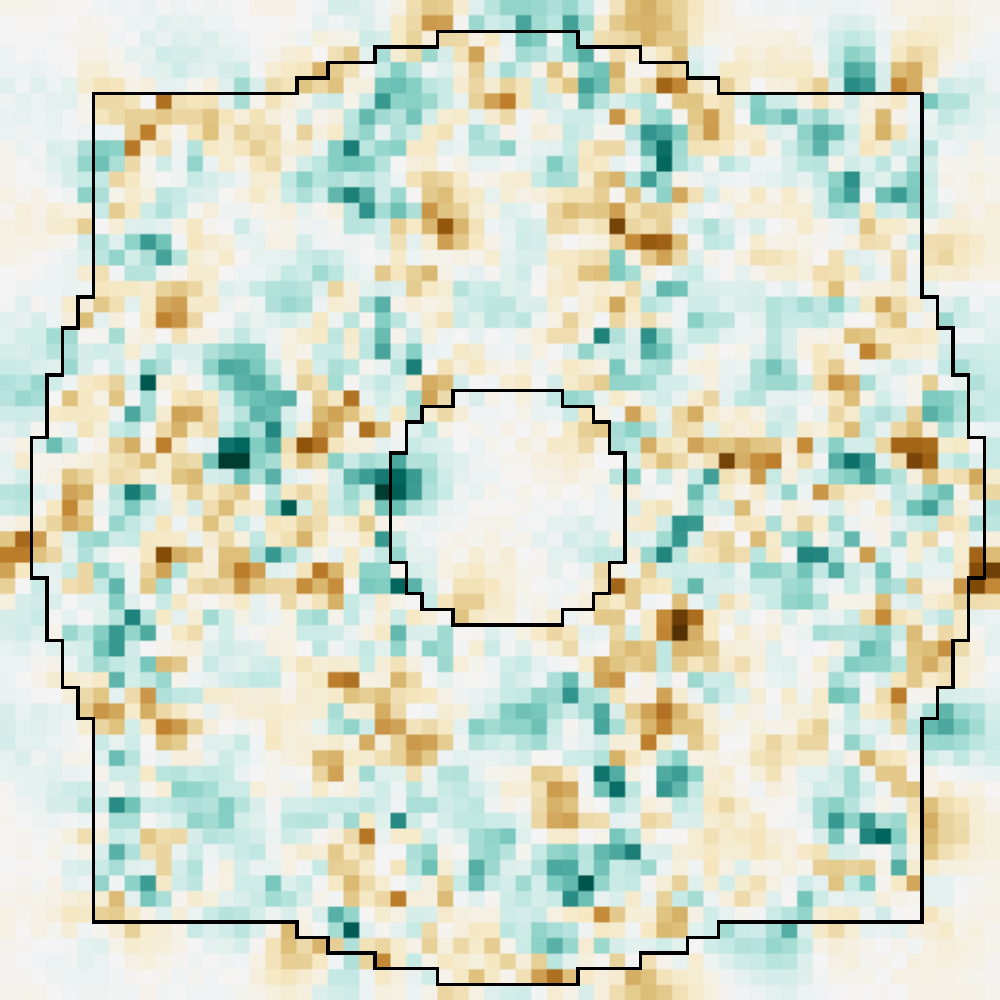}
\put(30,-10){\textsf{\scriptsize (d) Reconstructed Field }}
\end{overpic}
\vspace{1em}
\begin{overpic}[clip,trim={1cm 0.80cm 0 0cm},width=0.495\textwidth]{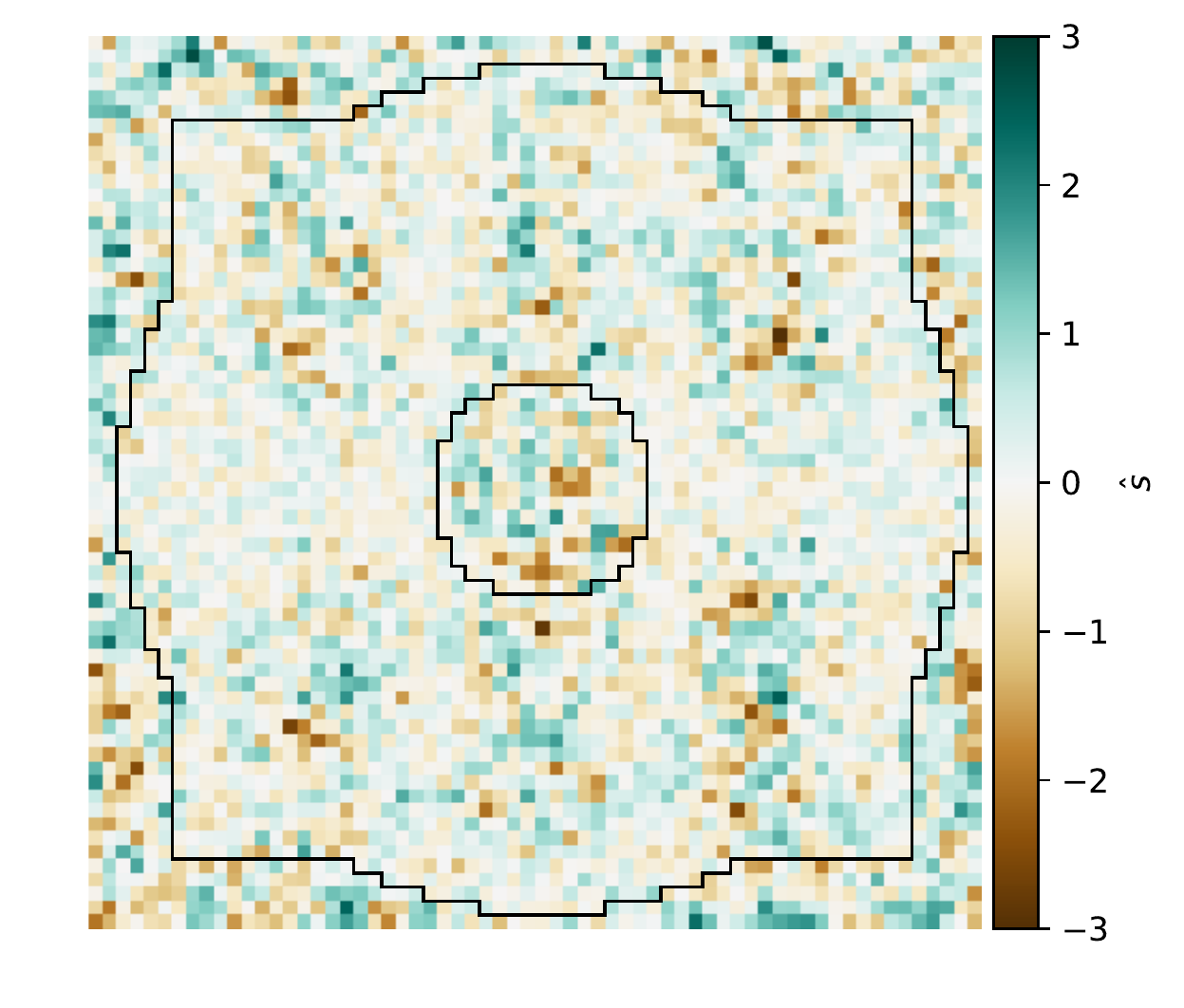}
\put(20,-10){\textsf{\scriptsize (e) True Field - Reconstructed }}
\end{overpic}
\end{center}

\caption{\label{fig:64_d}
MAP density reconstruction for the 64x64 case.
}
\end{figure}

\begin{figure}
\begin{center}
\includegraphics[width = 0.63\textwidth]{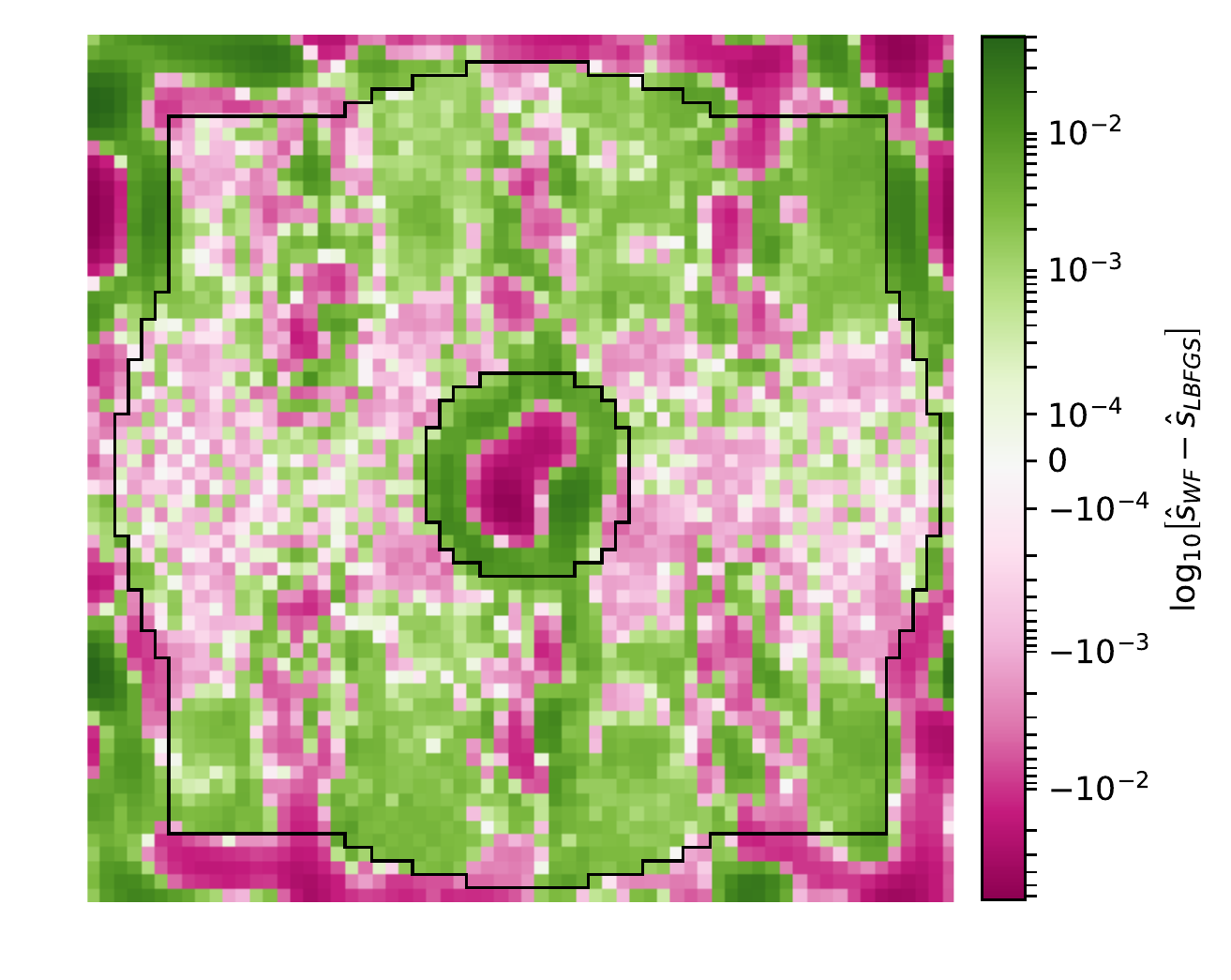}
\end{center}
\caption{\label{fig:64_wfvsml}
The log absolute magnitude difference of the direct matrix inversion Wiener filter solution and that attained via a MAP method. Note that differences are extremely small throughout the map and are particularly small in the unmasked region.
}
\vspace*{\floatsep}
\begin{center}
\includegraphics[width = 0.66\textwidth]{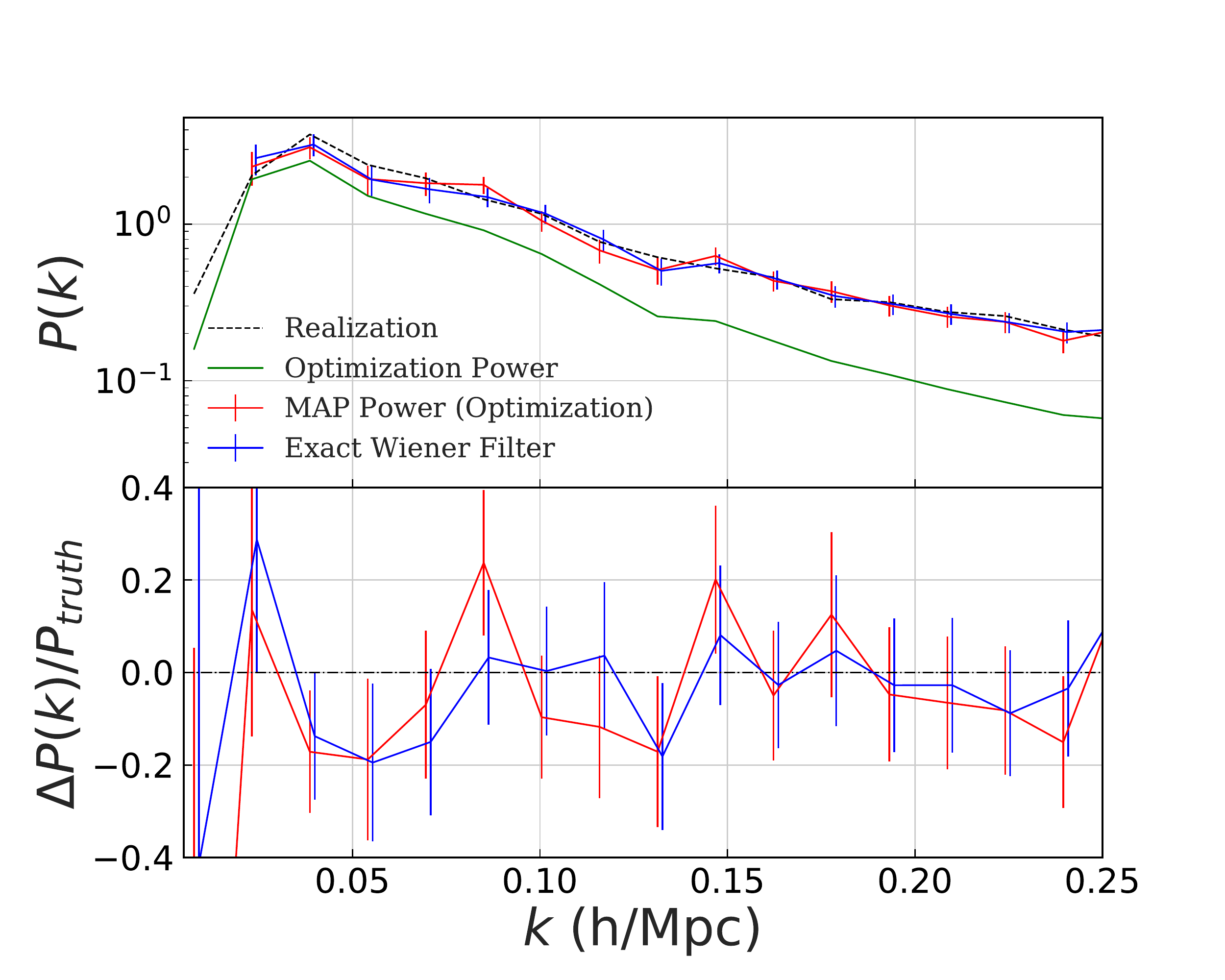}
\end{center}
\caption{\label{fig:pk64_wfvsLBFGS}Comparison of the maximum likelihood power spectrum attained from optimization versus that for brute force matrix inversion Wiener filter. Errors are visually identical, and all points $k$ bins are within one standard deviation of each other. 
}
\end{figure}

Using the input power spectrum of Eq. \ref{eq_simpleps} we generate a density field shown in Fig \ref{fig:64_d}(a), apply a mask and anisotropic noise shown in Fig \ref{fig:64_d}(b), which results in a mock observation in Fig \ref{fig:64_d}(c). We perform the minimization routine outlined in Sec. \ref{subsec:ml}, with the optimized map shown in Fig. \ref{fig:64_d}(d) with residuals shown in Fig \ref{fig:64_d}(e). There is qualitative agreement between the truth and the reconstructed field within the mask in the low-noise regions and is even able to reconstruct the larger scale modes right on the border within the masked region. However, as expected, the small scale modes within the high noise regions within the mask are poorly reconstructed since it is impossible to differentiate those modes in real space with the noise.

For this small test case we can compare the optimization result directly against a numerical inversion Wiener filter solution, which we show on the map level in Fig \ref{fig:64_wfvsml}. The results match outside the masked region within $10^{-4}$, while in the masked region there is a slightly large difference due to the imposition of a convergence criteria in our L-BFGS scheme. As we increase the required precision of the L-BFGS in terms of $\epsilon \equiv \Delta \chi^2$, we asymptotically approach the Wiener filter solution.

Using the formulation in Sec \ref{subsec:ps_est}, we can look at the performance of the technique as a function of scale. In Fourier space we can account for the reduction of small scale power caused by noise and also estimate the Hessian matrix (thereby giving error estimates). We show the power spectrum and error estimates from the optimization technique versus the direct Wiener filtering in Fig \ref{fig:pk64_wfvsLBFGS}. Note that the full reconstruction relies on both calculation of noise bias and Hessian matrix. We have compared each of these terms from the optimization method to those calculated via direct matrix inversion to confirm they are equal within the error of the required optimization precision. Also note that we only used one noise realization to estimate the noise bias. In general, the number of noise realizations necessary to appropriately estimate the number of underlying band-powers will depend on both the underlying noise model and the band-powers of interest. In this particular case we found the improvements from including multiple noise realizations minimal as the effect on the overall power-spectrum were sub $1 \%$. As with the map-level reconstruction, we find that decreasing the $\epsilon$ criteria leads to asymptotic convergence to the Wiener filtered solution.

\begin{figure}[t]

\begin{center}
\begin{overpic}[width=0.35\textwidth]{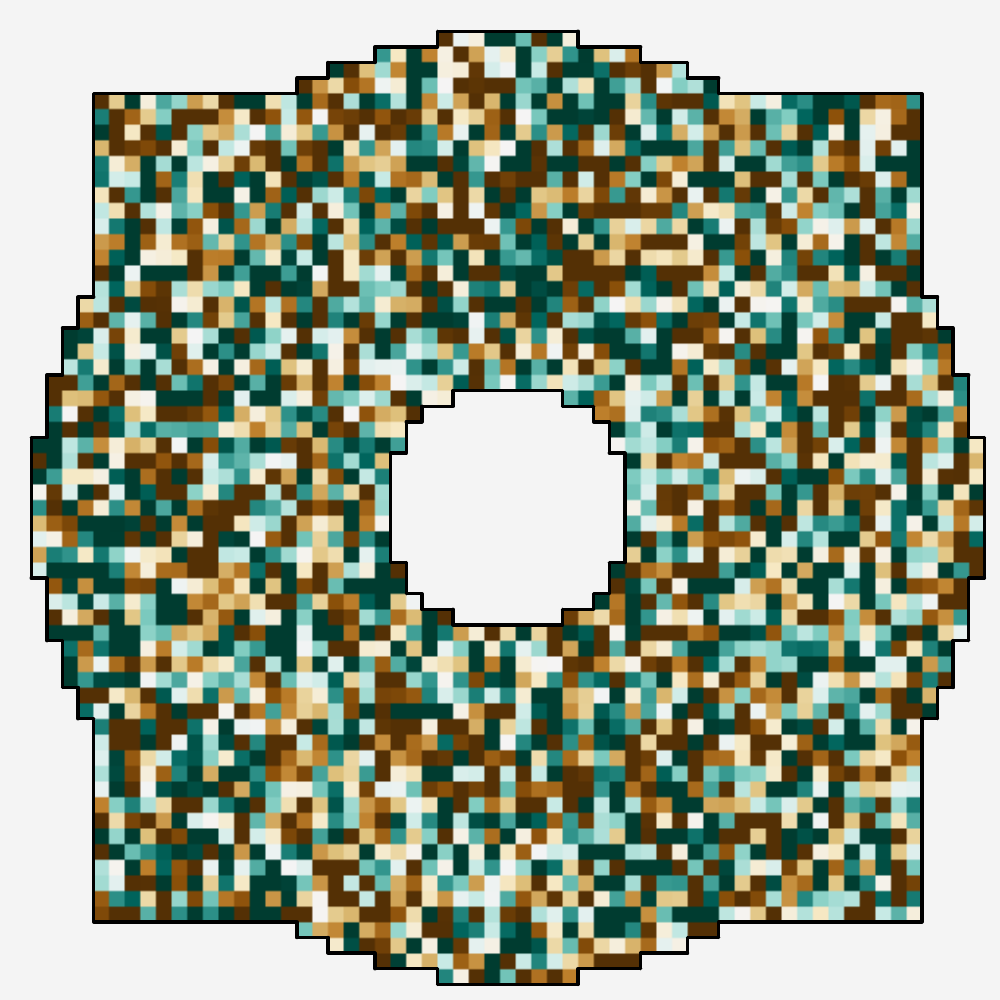}
\put(30,-10){\textsf{\scriptsize (a) High Noise Density Field }}
\end{overpic}
\vspace{1em}
\begin{overpic}[width=0.495\textwidth]{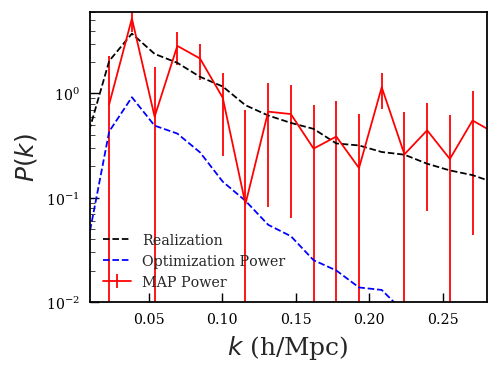}
\put(20,-10){\textsf{\scriptsize (b) Reconstructed Bandpowers}}
\end{overpic}
\end{center}

\caption{\label{fig:Noisey}
Observed field and associated reconstructed bandpowers for the noise dominated regime. Mask and color scale are the same as in Fig \ref{fig:64_wfvsml}, and noise is uniform over the field.
}
\end{figure}

We also explored the noise dominated regime more explicitly in Figure \ref{fig:Noisey}, where we apply a uniform high noise level over the entire field with variance 1.5 times the average variance in density. In this regime the optimized power is significantly suppressed across all scales, but the modes are still recoverable.

\subsection{Cosmic Microwave Background Temperature}
\label{subsec:CMBT}
\begin{figure}[t]
\begin{center}
\begin{overpic}[width=0.32\textwidth]{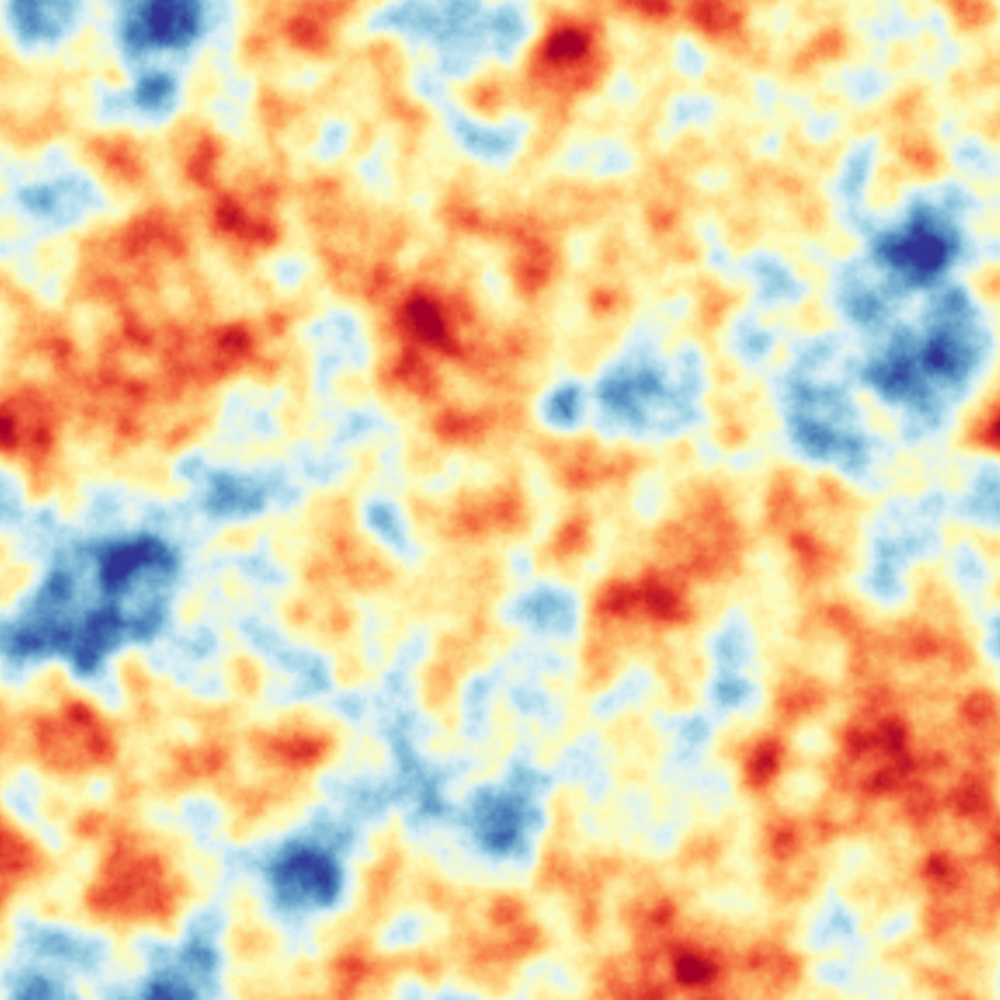}
\put(25,-10){\textsf{\scriptsize (a) Original Density Field }}
\end{overpic}
\begin{overpic}[width=0.32\textwidth]{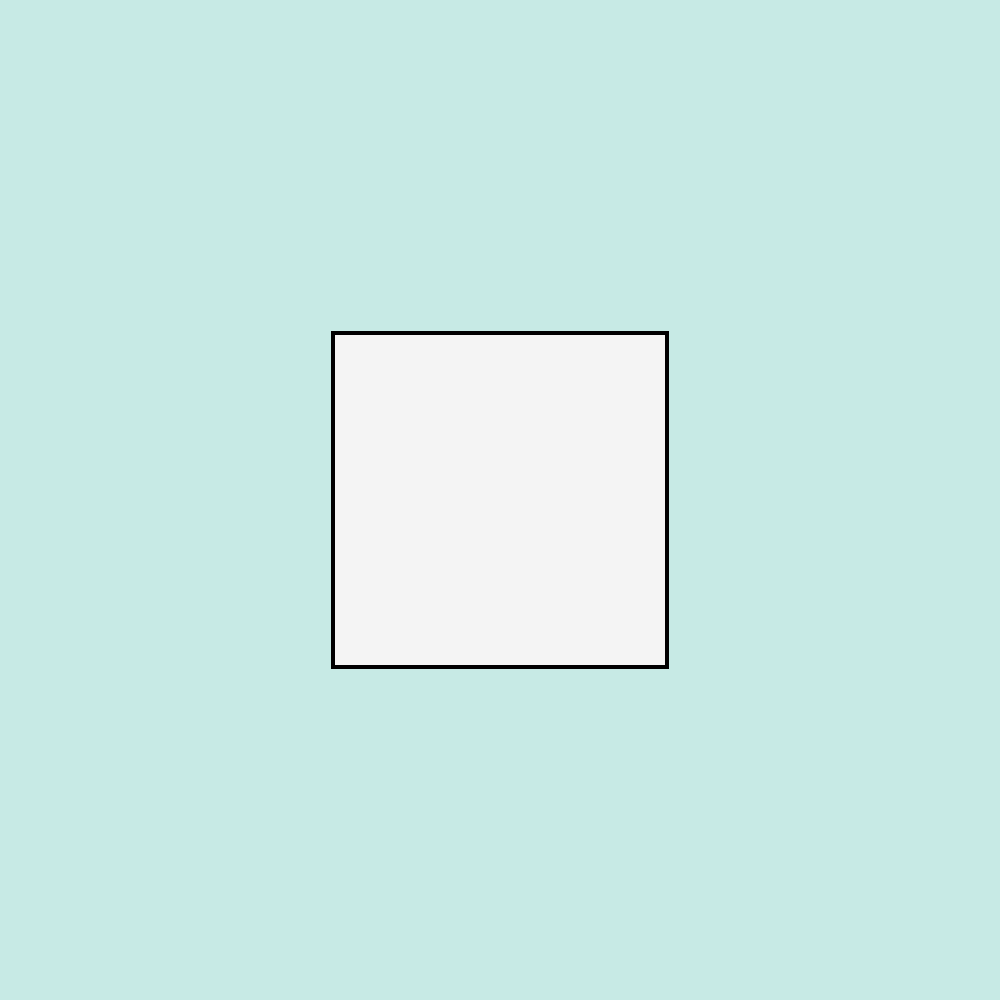}
\put(14,-10){\textsf{\scriptsize (b) Mask and Noise}}\end{overpic}
\vspace{1em}
\begin{overpic}[width=0.32\textwidth]{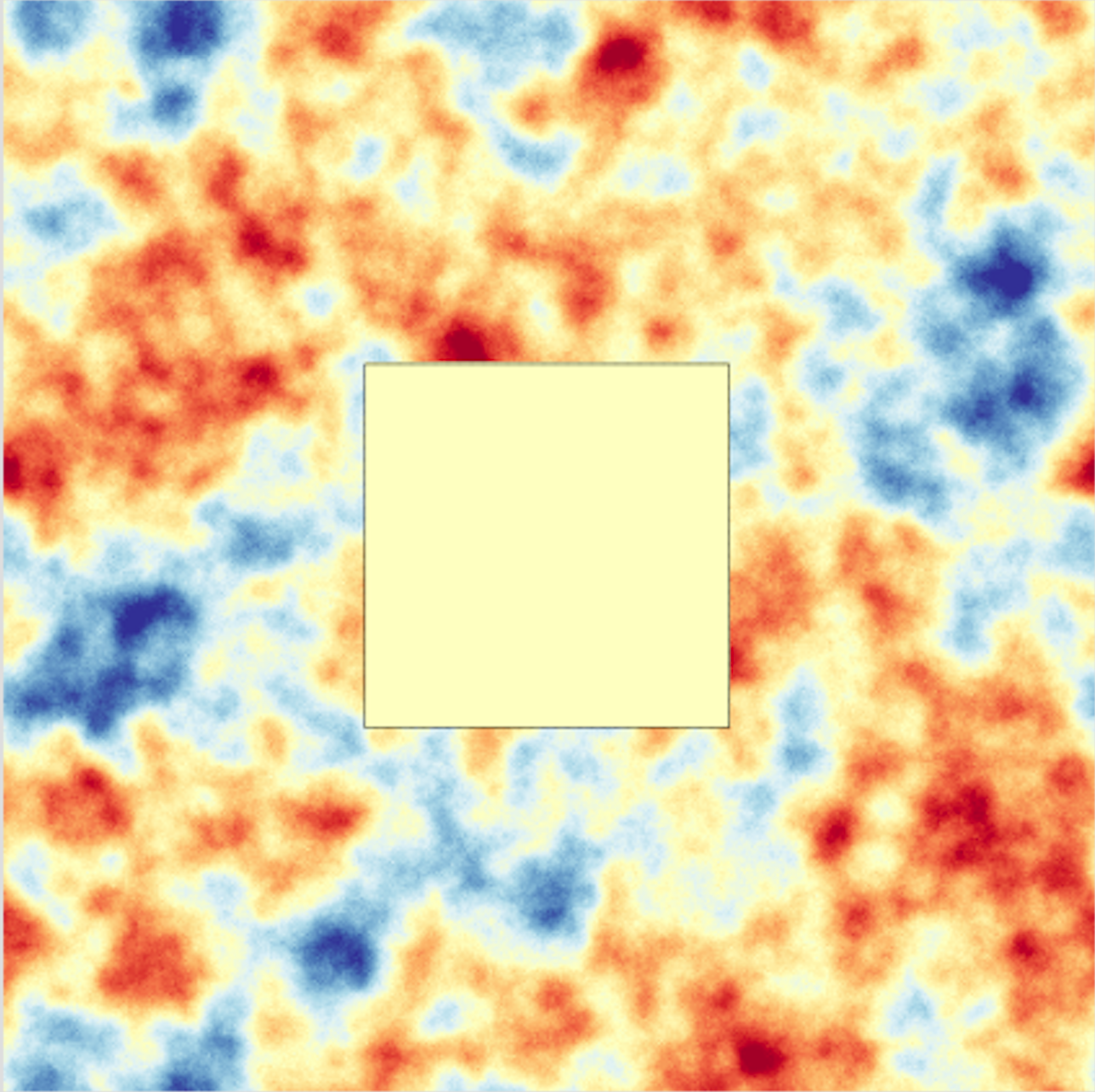}
\put(25,-10){\textsf{\scriptsize (c) Observed CMB Field}}
\end{overpic}
\end{center}
\vspace{-0.4cm}

\begin{center}
\begin{overpic}[width=0.41\textwidth]{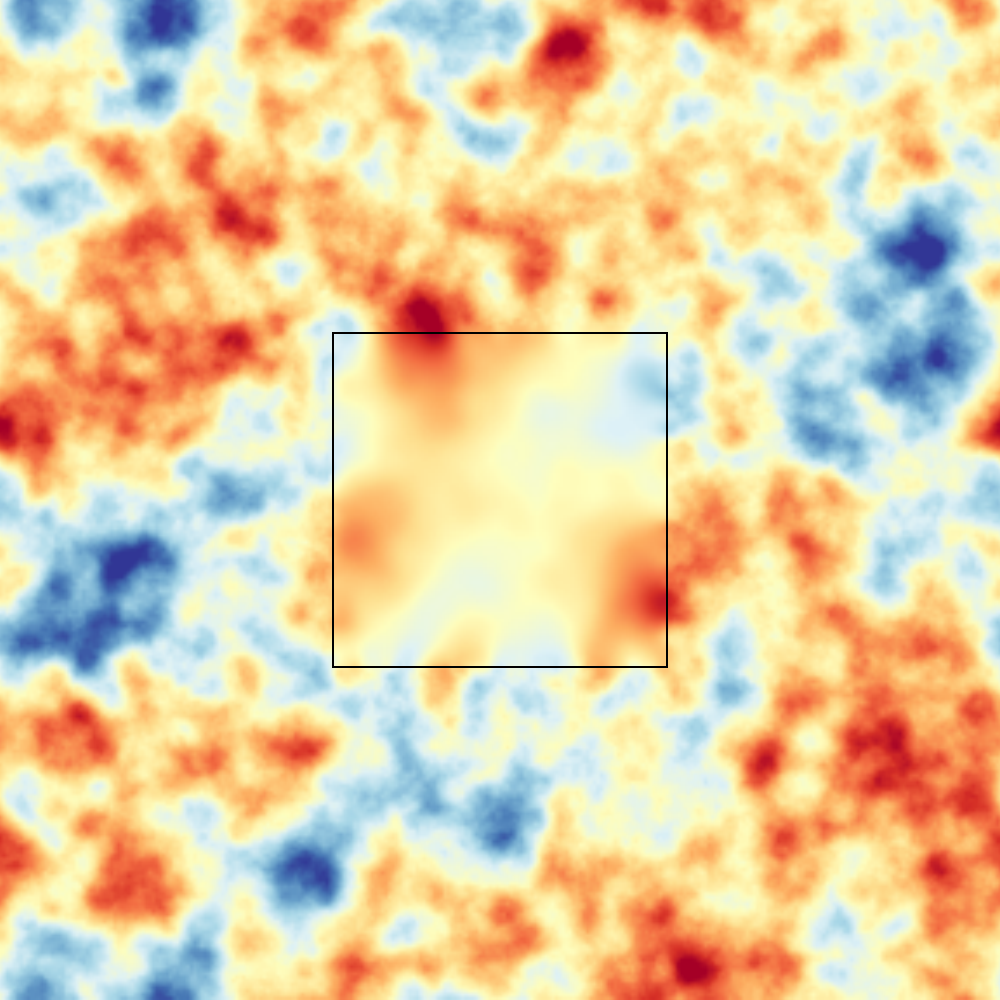}
\put(30,-10){\textsf{\scriptsize (d) Reconstructed Field }}
\end{overpic}
\vspace{1em}
\begin{overpic}[clip,trim={1cm 0.60cm 0 0cm},width=0.495\textwidth]{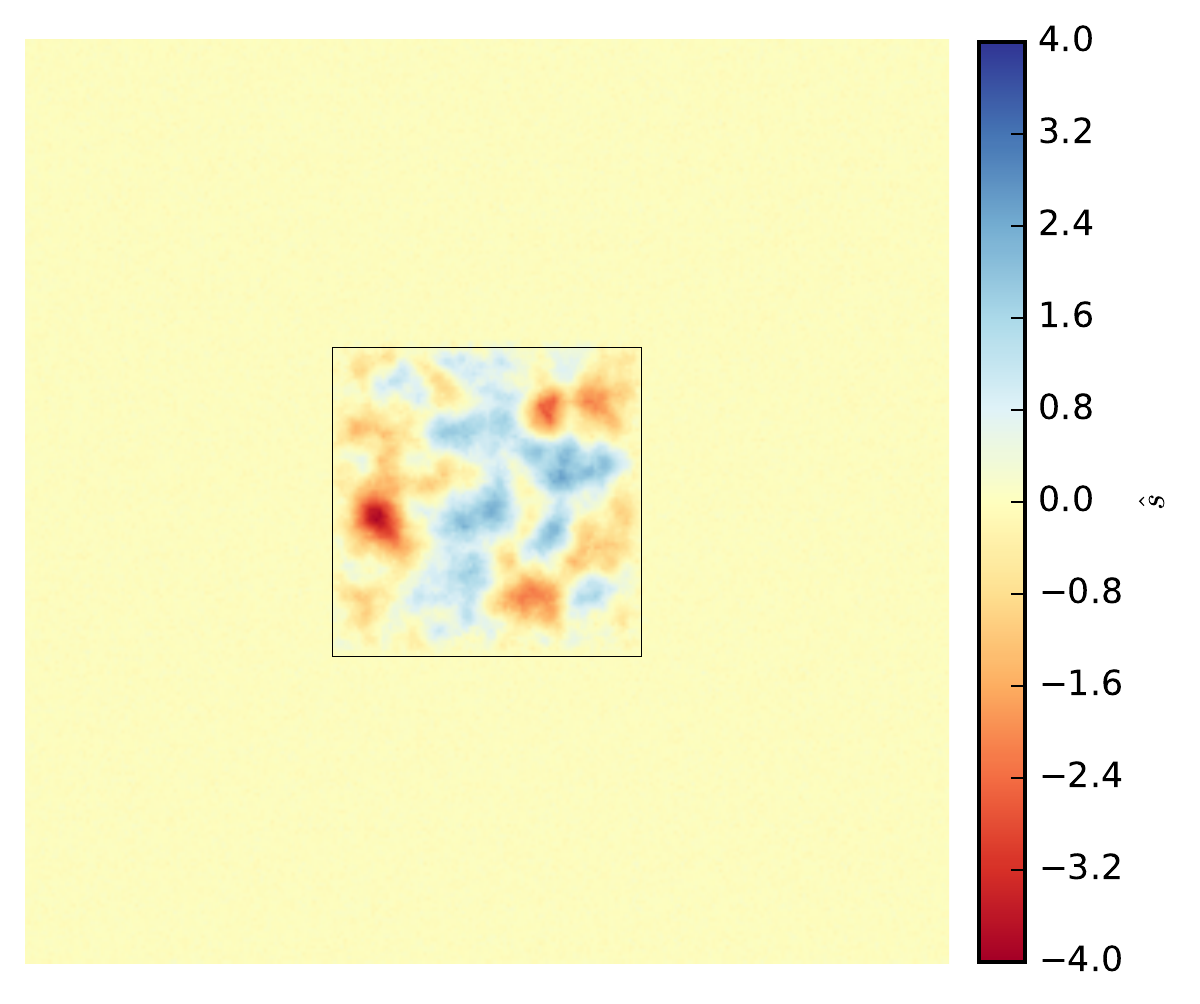}
\put(20,-10){\textsf{\scriptsize (e) True Field - Reconstructed }}
\end{overpic}
\end{center}

\caption{\label{fig:512CMB}
MAP CMB reconstruction for 512x512 pixel map. Note that images (a), (c), (d),and (e) have same absolute color scale, while (b) shows the spatial variance of the noise properties. Color scale is normalized to show standard deviations away from mean.
}
\end{figure}

The question of optimal reconstruction of CMB maps given irregular sky coverage and variable noise and foreground subtraction is a common issue for existing CMB surveys. So far, true maximum likelihood power spectra estimators have only been applied to data from WMAP \cite{Tegmark2003wmap} and for the largest angular scales in Planck data \cite{Planck15ps}, but these techniques are difficult to scale to the entire Planck dataset due to the significant increase in computational cost. 

While Planck's power spectrum measurements, and therefore cosmological parameter constraints \cite{Planck15}, do not rely on construction of the actual full map, other spatially dependent signals do. Cross-correlations between the primary CMB and other cosmological probes, such as x-ray signal or galaxy positions, require an accurate spatial reconstruction of the CMB map. In addition, full sky CMB lensing maps are constructed by applying the quadratic estimators to the CMB map (in either temperature, polarization, or some minimum variance combination of the two) and will similarly suffer if the reconstructed temperature and/or polarization maps are suboptimal.
In terms of map reconstruction, there are highly efficient Wiener filter programs available \cite{2017Seljebotn} which perform a multi-scale analysis with preconditioned optimization routines, which could replace our L-BFGS, but 
since the number of iterations needed is already low we did not explore this 
further. 

In Figure \ref{fig:512CMB}
we show results for temperature, without polarization. We mention that polarization 
would be analogous to the example of joint optimization of E and B fields, which we 
do in the case of cosmic shear in Appendix \ref{app:EB}.

The case of CMB reconstruction is analogous to that of the density field, but with a significantly redder spectrum. The condition number of the covariance 
matrix is thus significantly larger, and BFGS needs more iterations to converge. 
While we could have used conjugate gradient with a preconditioner (as in \cite{2017Seljebotn})  to improve the convergence
of the optimization step, we chose not to do so here since the computational cost was not 
significantly higher than that for the simple density case even for this case (see Figure \ref{comp}). 
For the implementation of our algorithm it is important to recognize that the increase in 
power on larger scales in the CMB case makes the masked region the most computationally expensive region to reconstruct, and a redder spectrum will allow more mode reconstruction within this region. For more discussion, see Appendix \ref{app:CMB_rec}.

We generate a mock primary CMB full-sky field using HEALPIX \cite{healpix} based on power spectrum generated from CLASS using the Planck 2015 cosmological parameters \cite{Planck15}. We then extract a $10 \times 10$ degree patch which we then mask a central region and introduce a white noise of 6 $\mu K$-arcmin. 

The reconstruction is quite good in the observed region as the presence of noise mostly affects small scales where there is very little power. In the masked region the optimization is able to reproduce some clear long-scale modes. In Appendix \ref{app:CMB_rec} we show that with a more accurate convergence criteria we reproduce more of these large scale modes in the masked region but in terms of power spectrum estimation this is unnecessary, as we are already cosmic variance limited even with $\epsilon = 0.1$.

\subsection{Cosmic Shear}
\label{subsec:cs}

\begin{figure}[t]
\begin{center}
\begin{overpic}[width=0.32\textwidth]{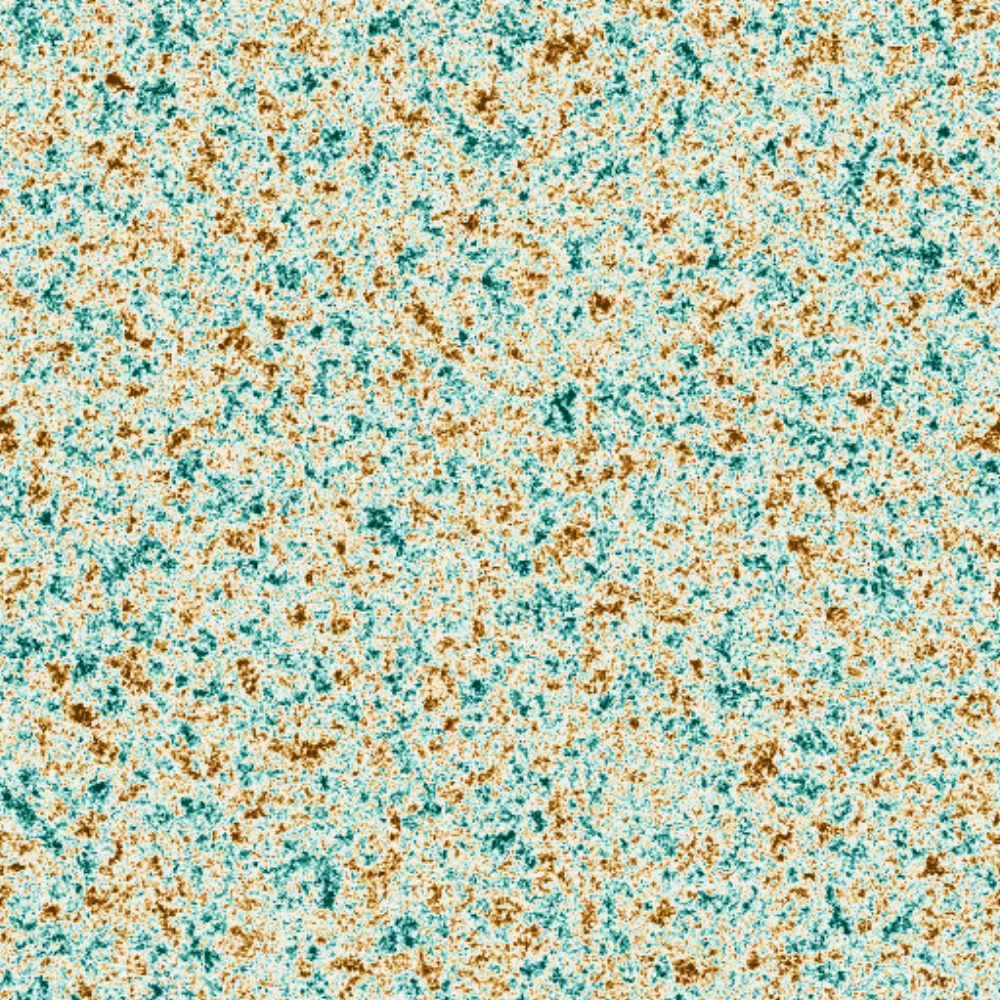}
\put(25,-10){\textsf{\scriptsize (a) Original Density Field }}
\end{overpic}
\begin{overpic}[width=0.32\textwidth]{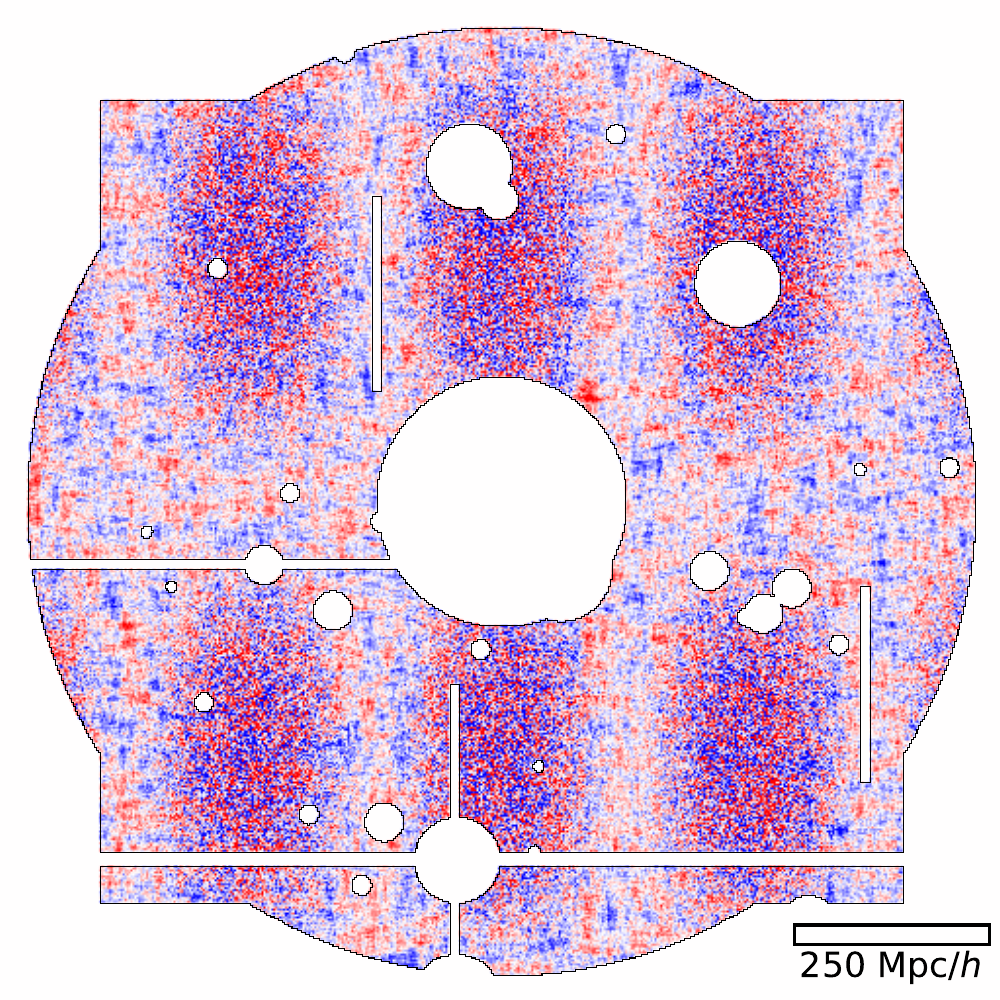}
\put(14,-10){\textsf{\scriptsize (b) Observed Shear Field, $\gamma_1$}}\end{overpic}
\vspace{1em}
\begin{overpic}[width=0.32\textwidth]{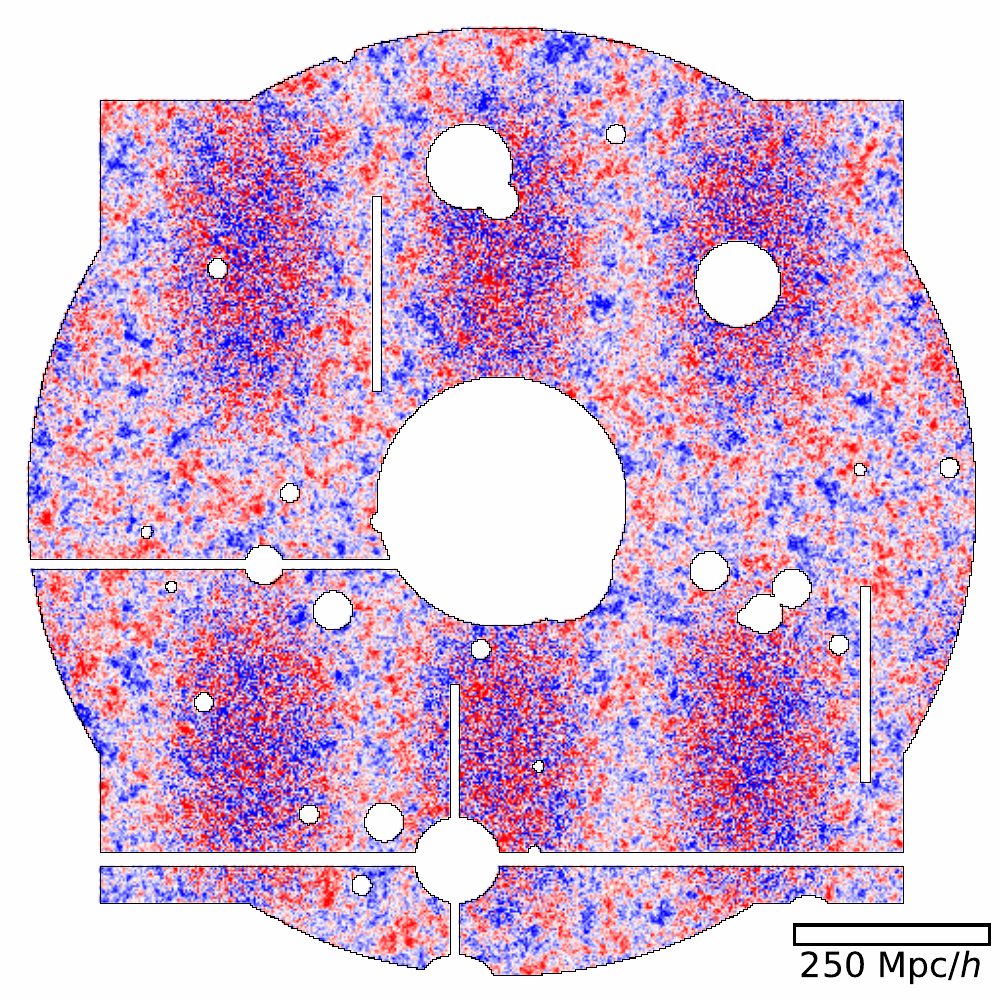}
\put(25,-10){\textsf{\scriptsize (c) Observed Shear Field, $\gamma_2$ }}
\end{overpic}
\end{center}
\vspace{-0.4cm}

\begin{center}
\begin{overpic}[width=0.495\textwidth]{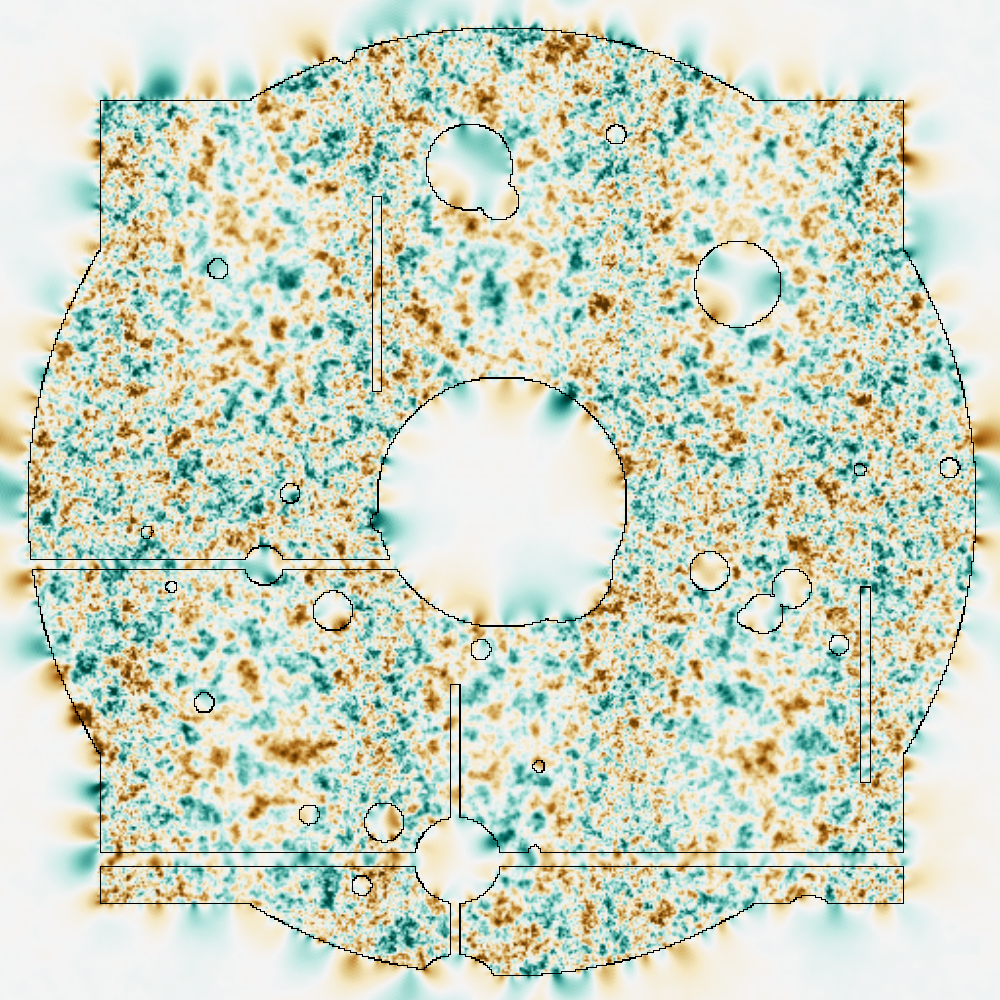}
\put(30,-10){\textsf{\scriptsize (d) Reconstructed Field }}
\end{overpic}
\vspace{1em}
\begin{overpic}[width=0.495\textwidth]{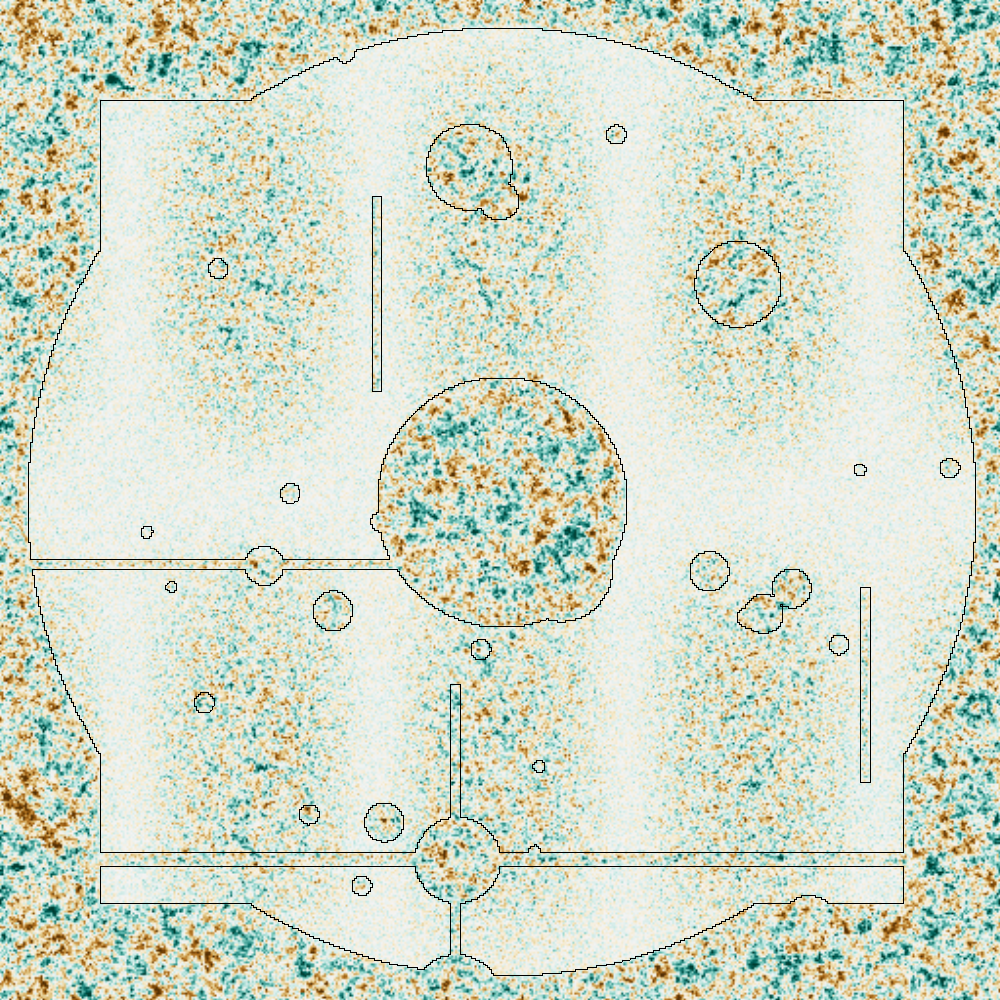}
\put(20,-10){\textsf{\scriptsize (e) True Field - Reconstructed }}
\end{overpic}
\end{center}

\caption{\label{fig:512s_d}
Maximum likelihood shear reconstruction for a 512x512 pixel map. Here we use the same mask and noise properties, as well as color scaling, as shown in Figure \ref{fig:512_d}. 
}
\end{figure}

\begin{figure}
\begin{center}
\includegraphics[width = 0.8\textwidth]{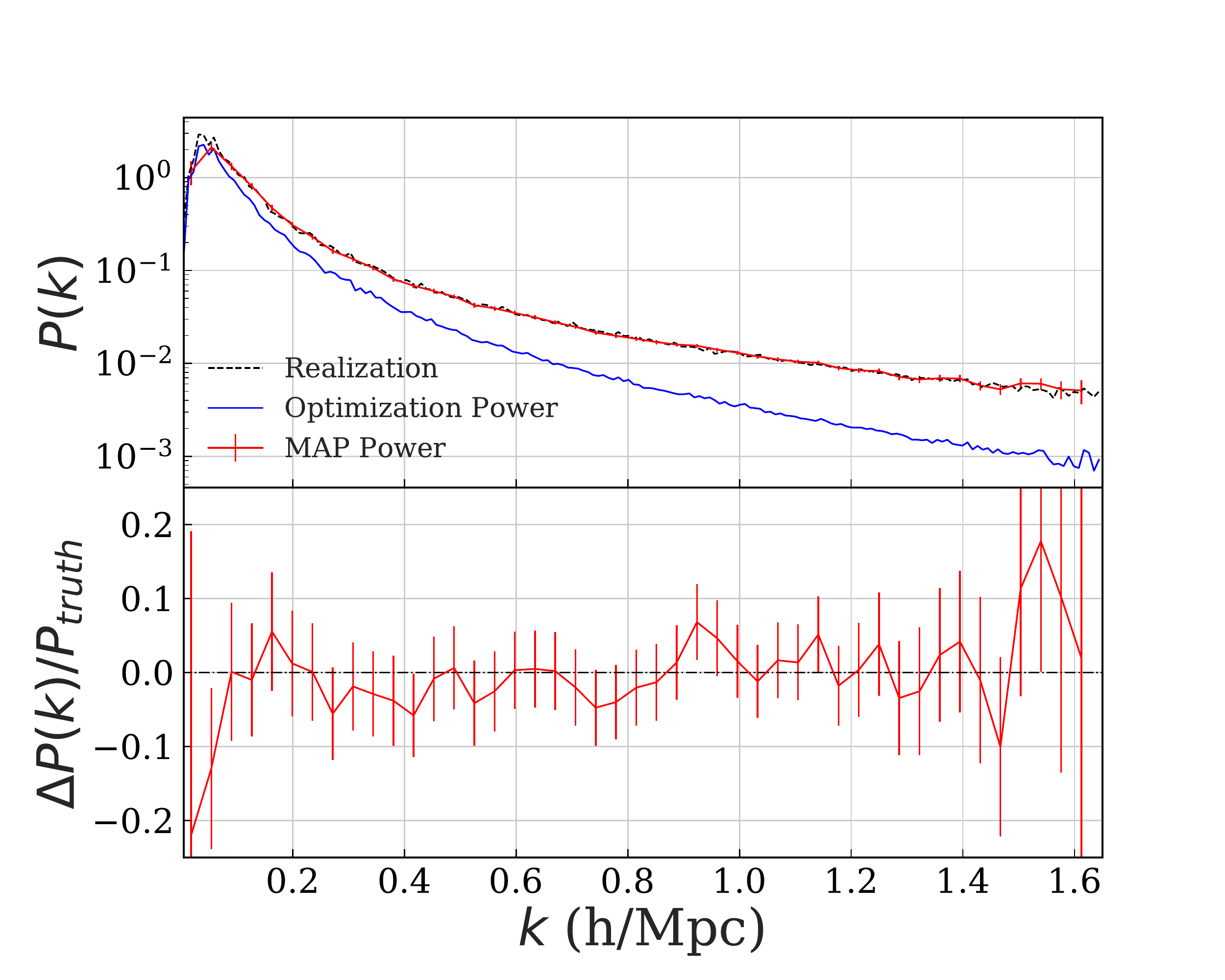}
\end{center}
\caption{\label{fig:Shearpk512}Comparison of the maximum likelihood power spectrum attained from optimization versus the true power-spectrum of the region for the shear-only reconstruction. Also shown is the importance of the noise bias correction (or, equivalently, the importance of the Hessian determinant). 
}
\end{figure}

In this section we specialize to only fitting a curl-free $E$ component; we discuss the more general case including a curl component in Appendix \ref{app:EB}. For a details on the cosmic shear formalism, see \cite{van2001cosmic,bartelmann2001weak,jain1997cosmological}.

To apply our method, we perform optimization over the underlying density field and at each step of the optimization compute the corresponding shear maps, $\gamma_1$ and $\gamma_2$, to compare with the mock observed shear maps. In principle, instead of working with the shear maps which require binning and/or interpolation between galaxies, one could work directly with the catalog of galaxy shapes and compute the likelihood of the observed ellipticity instead of the given averaged shear maps (as in \cite{PadmanabhanSeljakEtAl03,Bohm2017}). 

This method is in contrast to the standard Kaiser-Squires (KS) \cite{KaiserSquiresEtAl95} technique which has proven quite successful so far in cosmic shear analysis and cluster mass estimation. However, KS has some notable downsides in the presence of anisotropic noise or a mask as it is not able to self consistently down-weight the high noise areas and masked regions, resulting in defects on boundaries. In addition, the noise inherent in these measurements propagates onto the final mass-maps, resulting in an inaccurate small scale power measurement. This has been shown to be particularly detrimental to peak statistics measurements \cite{ImprovingWLMM2018}.

The results in position space are shown in Figure \ref{fig:512s_d}. In Figure \ref{fig:512s_d}(a) we show the initial density field and in Figure \ref{fig:512s_d}(b,c) we show the observed shear fields including mask and noise properties. Our reconstructed maximum likelihood map is shown in Figure \ref{fig:512s_d}(d) and the difference between the original field and reconstructed in Figure \ref{fig:512s_d}(e). As in the case of the density field, the optimization technique is able to exactly reconstruct the density in the low noise, unmasked regions, but only recovers large scale scale power in the higher noise unmasked regions.

\section{Discussion and Conclusion}
\label{sec:conclusions}

In this work we have demonstrated that is is possible to efficiently reconstruct the MAP signal field and the maximum likelihood power spectrum for linear fields for realistic survey sizes. This technique is equivalent to the Wiener filter solution for small enough convergence criteria and has been applied to a number of cosmological fields (density, CMB, and cosmic shear maps). We are able to reconstruct the initial density field and the overall power spectrum, accounting for noise bias and window-function effects due to masking. We first recast the field reconstruction into an optimization problem, which we solve using quasi-Newton optimization. 
We then recast the power spectrum estimation into the field marginalization problem, from which we obtain an 
expression that depends on the field reconstruction solution and a determinant term. We develop a novel simulation based 
method for the latter. We extend the simulations formalism to provide the covariance matrix for the power spectrum. 

\begin{figure}
\begin{center}
\includegraphics[width = 0.65\textwidth]{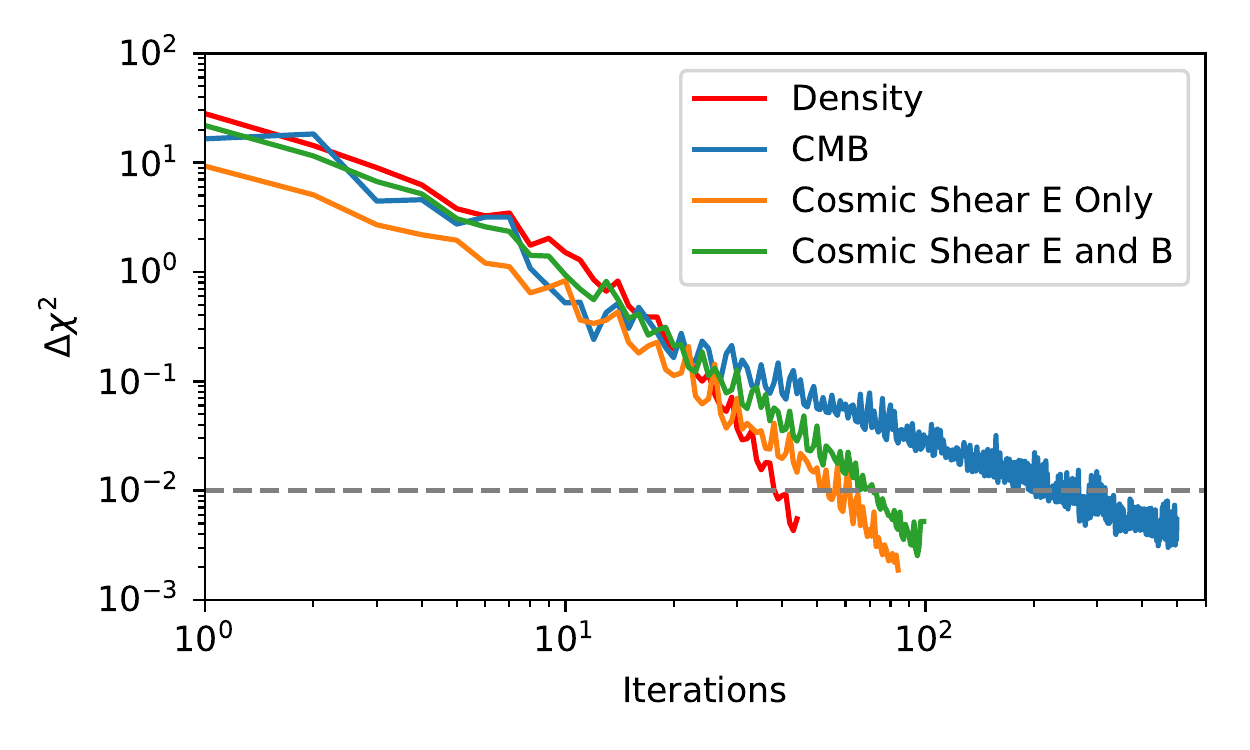}
\end{center}
\caption{\label{fig:chi_all} Comparison of the convergence properties of the various cosmological density fields studied in this work. All cases have $512^2$ pixels, and comparable effective volume.
\label{comp}
}
\end{figure}

This technique outperforms the brute force Wiener filter technique in terms of computational time and memory requirements. True Wiener filter requires an inversion of the full pixel covariance matrix, $\bi{C}$, which for realistic surveys would be highly non trivial. Numerical methods approximate $\bi{C}^{-1}\bi{d}$, which allows for map-level reconstruction but by itself doesn't allow calculate of the Hessian matrix for band power reconstruction. This reconstruction also requires evaluation of the determinant of the Hessian, or its derivative, trace, and where the techniques provided in this work allows orders of magnitude improvement in realistic cases. We evaluate this determinant derivative using forward model realizations and additional optimization. This allows us to use off the shelf optimization codes such as L-BFGS, as well as convergence criteria to find the proper trade-off point between accuracy and computation time. We compare the two 
in Appendix \ref{subsec:wf}, finding good agreement for low dimensionality problems where brute force 
approach is feasible. 

In Figure \ref{fig:chi_all} we show the convergence properties for the 3 cases studied in Section \ref{sec:examples}, as well as the joint E/B cosmic shear case presented in Appendix \ref{app:EB}. While all the cases have a comparable number of unmasked pixels, convergence properties differ due to the difference of the underlying fields power-spectrum. More power at larger scales (i.e. a redder spectra) requires additional iterations to reconstruct the power within the masked regions.

While we used L-BFGS due to its well established optimization properties in very high dimensional convex optimization problems, we do not make a claim of optimality in terms of the particular technique for performing the optimization. L-BFGS constructs low rank Hessian approximation to the Hessian, which 
makes it a quasi-Newton method: the closer this Hessian is to the true Hessian 
the closer we are to true second order optimization. In this limit this method
will outperform any other method, including preconditioned conjugate gradient, 
which is only effective if preconditioning reduces the condition number of 
the problem. On the other hand, true second order Newton works for any 
condition number. We also note that 
since linear problems are convex,
an optimizer is always guaranteed to find the global minimum.
In general, we did not find particularly large performance changes when using other optimization techniques, such as conjugate gradient. Sampling based methods, like Hamiltonian Monte Carlo, are unnecessary for these linear cases, as there is no need to sample the distribution which is well approximated as a multivariate Gaussian, except for modes on order the size of the survey volume which are poorly sampled (and therefore do not follow the central limit theorem), in which case one can use approximations of inverse Wishart distribution developed 
in appendix A of \cite{seljak2017towards}.

A comparison can be made in our primary CMB example in Section \ref{subsec:CMBT} to the results of the messenger and dual messenger field found in \cite{KodiLavauxEtAl2017} (see Figure 6 in \cite{KodiLavauxEtAl2017}). The L-BFGS approach requires significantly fewer iterations than the messenger field ($\sim$ 5x) and dual messenger field ($\sim$ 2x) for a comparable convergence criteria. It is possible that an optimized cooling scheme for the messenger/dual messenger field would yield similar convergence properties, but this choice would likely be problem-specific and introduce an additional parameter to tune in the optimization. 

Going forward, it will be useful to extend this technique to other cosmological observables such as cosmic shear tomography (such as in \cite{2009simon}), Lyman - $\alpha$ tomography, and CMB lensing. Already work has been done applying this maximum likelihood approach to CMB lensing \cite{HirataSeljak03recon,CarronLewis17}, and further extending this work with these methods to small scales where standard quadratic estimators \cite{HuOkamoto02Estimator} are known to be suboptimal \cite{HorowitzFerraroSherwin17} would be promising future approach. Another avenue of particular interest is the ability for this technique to be useful for combining multiple (biased) tracers of some underlying field to create a maximum likelihood estimate of the field. One particularly promising example is jointly maximizing the underlying density likelihood function with respect to both the shear map as well as the projected galaxy density map \cite{2014Szepietowski,2013simon}.

These linear methods have a limitation when applied to nonlinear fields. 
Recent work \cite{ImprovingWLMM2018} has demonstrated that Wiener Filtering is not optimal in terms of detecting peaks in the density field and that sparsity-based reconstruction methods can yield higher signal to noise. More general reconstruction of density fields was explored in \cite{2018sparse}, where they created MAP estimates of physical clusters with associated error estimates without making assumptions about the Gaussian of the likelihood surface. However, these approaches are ad-hoc, 
as the loss function they minimize cannot be theoretically justified. 
For the case of nonlinear large scale structure, best analog of WF 
reconstructions in terms of minimizing the error are  
the nonlinear density reconstruction techniques developed e.g. in \cite{seljak2017towards,ECULID1,2018Jasche}, which first give a minimal variance reconstruction of initial Gaussian density 
fluctuations, and then project these into the nonlinear structures using 
an N-body simulation.

\acknowledgments
We would like to thank Vanessa B{\"o}hm, Simone Ferraro, Stephanie Ger, Chirag Modi, and Michael Schneider
for useful discussions and/or comments during the preparation of this manuscript. BH is supported by the NSF Graduate Research Fellowship, award number DGE 1106400. US acknowledges support from NSF 1814370, NSF 1839217, NASA 17-ATP17-0007 and NNX15AL17G. 

This research has made use of NASA's Astrophysics Data System.
This research used resources of the National Energy Research Scientific Computing Center, a DOE Office of Science User Facility supported by the Office of Science of the U.S. Department of Energy under Contract No. DE-AC02-05CH11231.

\appendix

\section{Wiener Filter Review}
\label{app:wf}
Wiener filtering (WF) \cite{wiener1949extrapolation, rybicki1992interpolation} is a popular way to non-parametrically reconstruct cosmological data as, in the linear case, it should minimize variance. In the absence of non-Gaussian sources of signal or noise, WF is optimal in the sense that it is equal to the maximum posterior probability estimator \cite{seljak2017towards}. Here we want to reconstruct the field itself, $\bi{s}$, given the noisy and/or poorly sampled data, $\bi{d}$. We define our estimated field $\bi{\hat{s}}=\bi{\Phi}\bi{d}$, where $\bi{\Phi}$ is a linear operator, i.e. a N x M dimensional matrix transforming from ``image space'' to ``field space''.  This can be found by attempting to minimize the variance of the residual, i.e. 
$\langle (\bi{s}-\bi{\hat{s}})(\bi{s}-\bi{\hat{s}})^\dagger \rangle$,
with respect to $\bi{\Phi}$. The Wiener filtered estimator is
\begin{equation}
\bi{\hat{s}} = \bi{\Phi}\bi{d}= \langle \bi{s d^\dagger}\rangle  \langle \bi{d d^\dagger}\rangle^{-1} \bi{d} = \bi{S} \bi{R^\dagger}\bi{C}^{-1}\bi{d},
\label{eq:wiener}
\end{equation}
which will result in a variance of residuals of the form
\begin{equation}
\langle (\bi{s}-\bi{\hat{s}})(\bi{s}-\bi{\hat{s}})^\dagger \rangle = \bi{S} - \bi{SR}^\dagger\bi{C}^{-1}\bi{RS}.
\end{equation}

Wiener filter only uses the mean and variance of the statistical distribution. If our underlying field is strongly non-Gaussian the WF may no longer be optimal. However, it will still minimize the variance, as defined in equation \ref{eq:var}, just that this minimization of variance may not correspond to the notion of the best reconstruction as it only captures the two point statistics of the underlying field. It may be  difficult to even define a measure to use for optimally of reconstruction in these cases, although nonlinear reconstruction methods certainly exist \cite{seljak2017towards}.

We now want to connect the Wiener Filter solution to the optimal power spectrum estimator. We multiply equation \ref{eq:ql} by the Hessian matrix,
\begin{eqnarray}
(\bi{F}\Theta)_l =& \frac{\bi{F}}{2}\sum_{l'}F^{-1}_{ll'}(\bi{d}^\dagger \bi{C}^{-1} \bi{Q}_{l'} \bi{C}^{-1}\bi{d} - b_{l'}),\nonumber \\
=& \frac{\delta_{ll'}}{2}(\bi{d}^\dagger \bi{C}^{-1} \bi{\Pi}_{l'} \bi{R} \bi{R}^\dagger \bi{\Pi}_{l'} \bi{C}^{-1}\bi{d} - b_{l'}).
\end{eqnarray}
The noise $\bi{b}$ can be similarly transformed as 
\begin{equation}
b_l = \text{tr} (\bi{\Pi}_l\bi{R}^\dagger \bi{C}^{-1}(\bi{N})\bi{C}^{-1}\bi{R}\bi{\Pi}_l),
\end{equation}
and the Hessian matrix itself as
\begin{equation}
F_{ll'} = \text{tr} (\bi{C}^{-1} \bi{Q}_l\bi{C}^{-1}\bi{Q}_{l'}) = \frac{1}{2} |\bi{\Pi}_l\bi{R}^\dagger \bi{C}^{-1}\bi{R}\bi{\Pi}_l|^2.
\label{eq:wf_exact_f}
\end{equation}
Both the Wiener Filter and the optimal power spectrum estimator first weigh the data by the inverse covariance matrix, essentially down weighting modes that either have high measurement error or strong correlation with other measurements. 





\section{Convergence Criteria of CMB Reconstruction}
\label{app:CMB_rec}

\begin{figure}
\begin{center}
\begin{overpic}[width=0.32\textwidth]{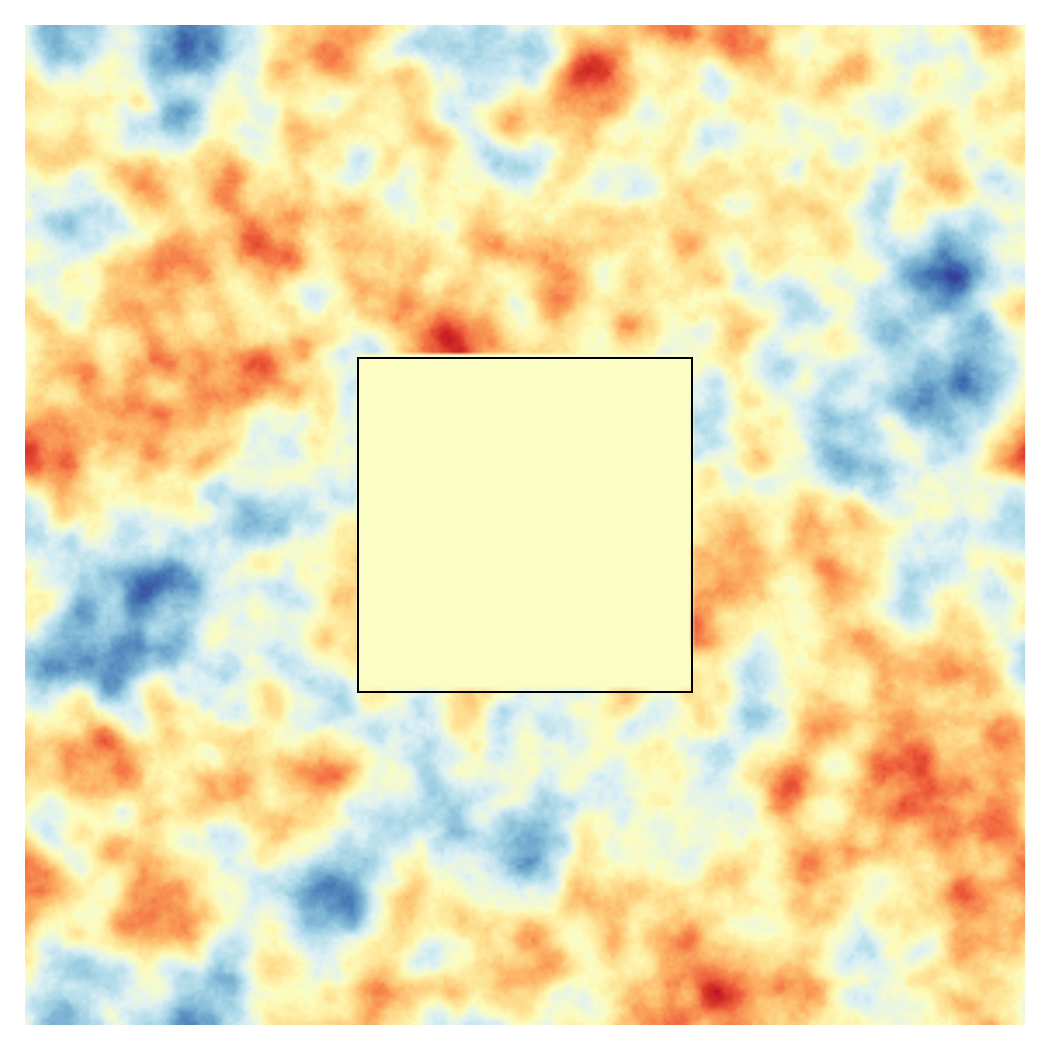}
\put(25,-10){\textsf{\scriptsize (a) $i = 1$}}
\end{overpic}
\begin{overpic}[width=0.32\textwidth]{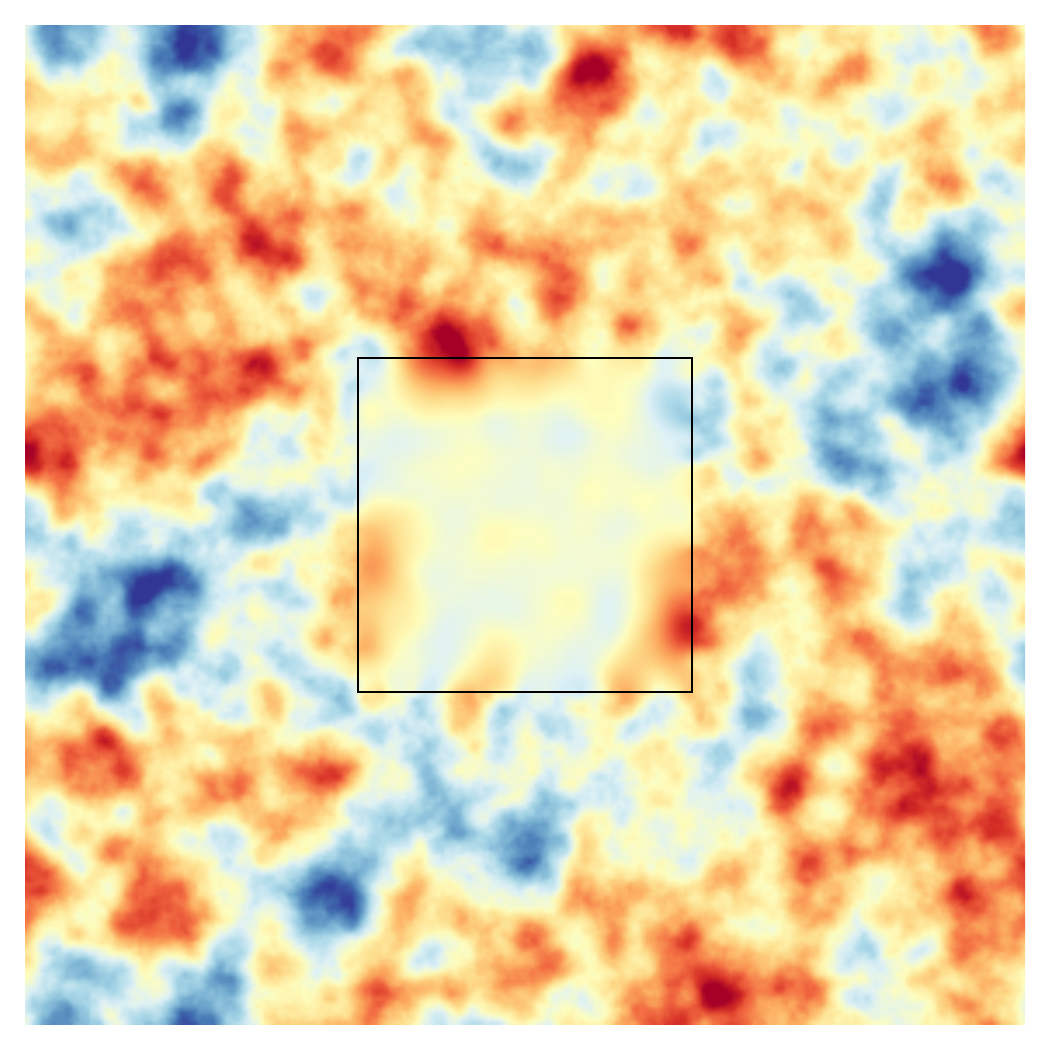}
\put(25,-10){\textsf{\scriptsize (b) $i = 1000$}}\end{overpic}
\vspace{1em}
\begin{overpic}[width=0.32\textwidth]{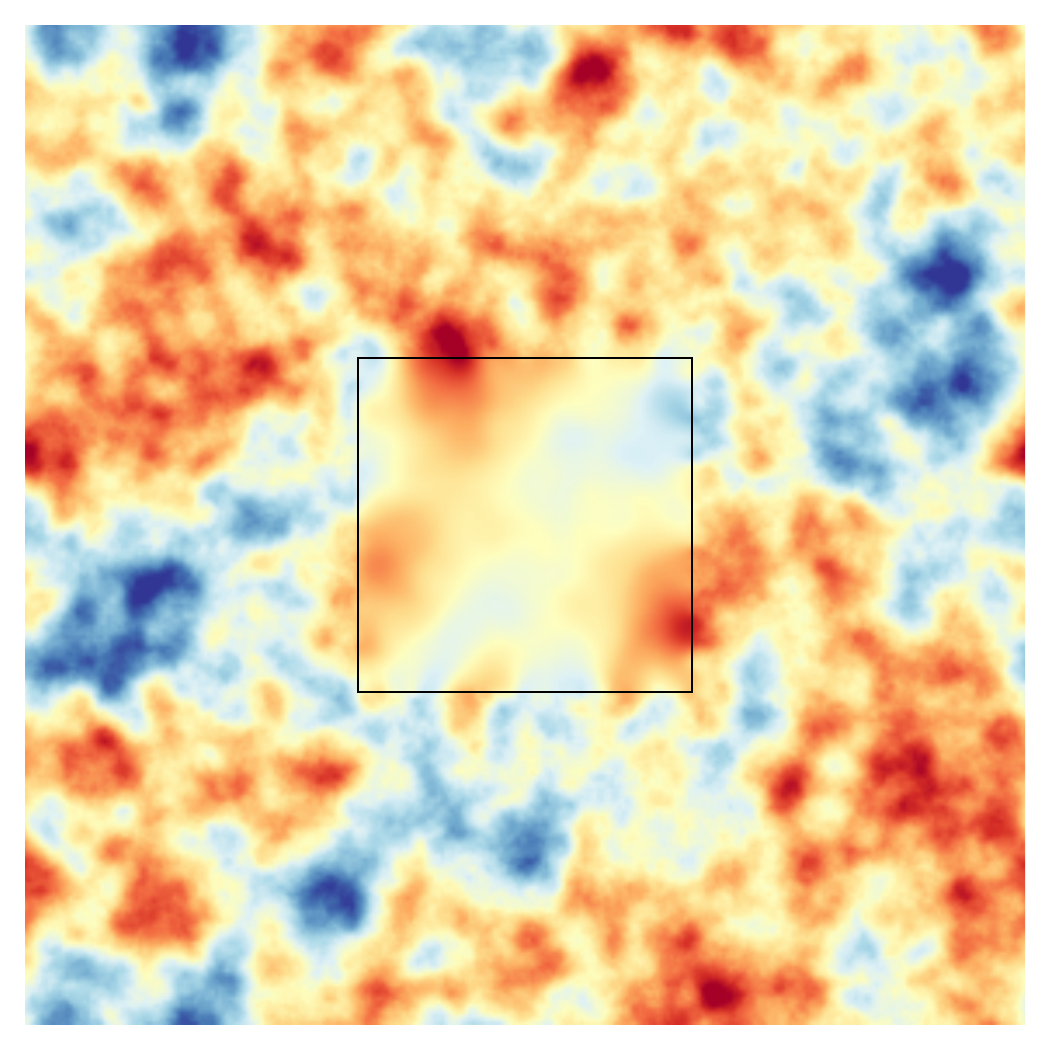}
\put(25,-10){\textsf{\scriptsize (c) $i = 5000$}}
\end{overpic}
\end{center}
\vspace{-0.6cm}
\begin{center}
\begin{overpic}[width=0.32\textwidth]{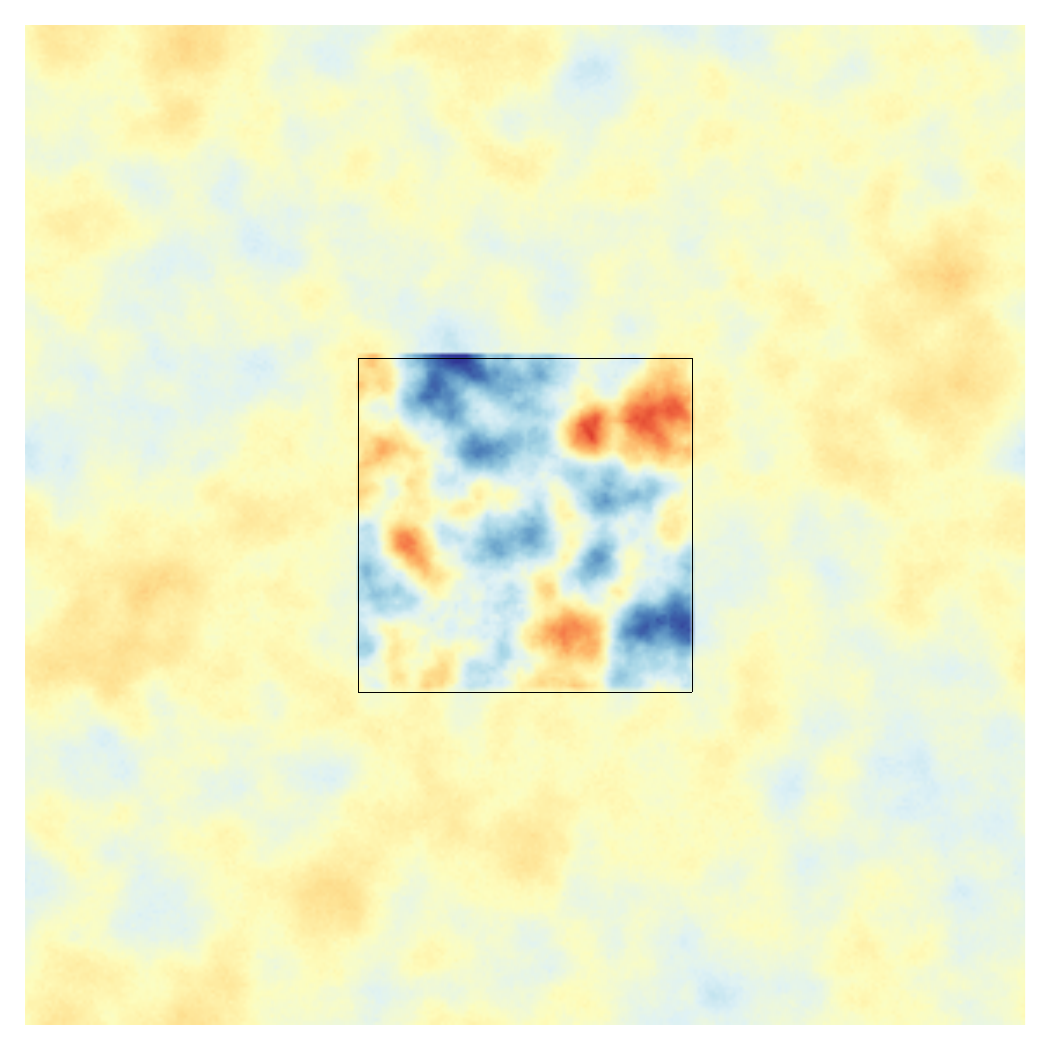}
\end{overpic}
\begin{overpic}[width=0.32\textwidth]{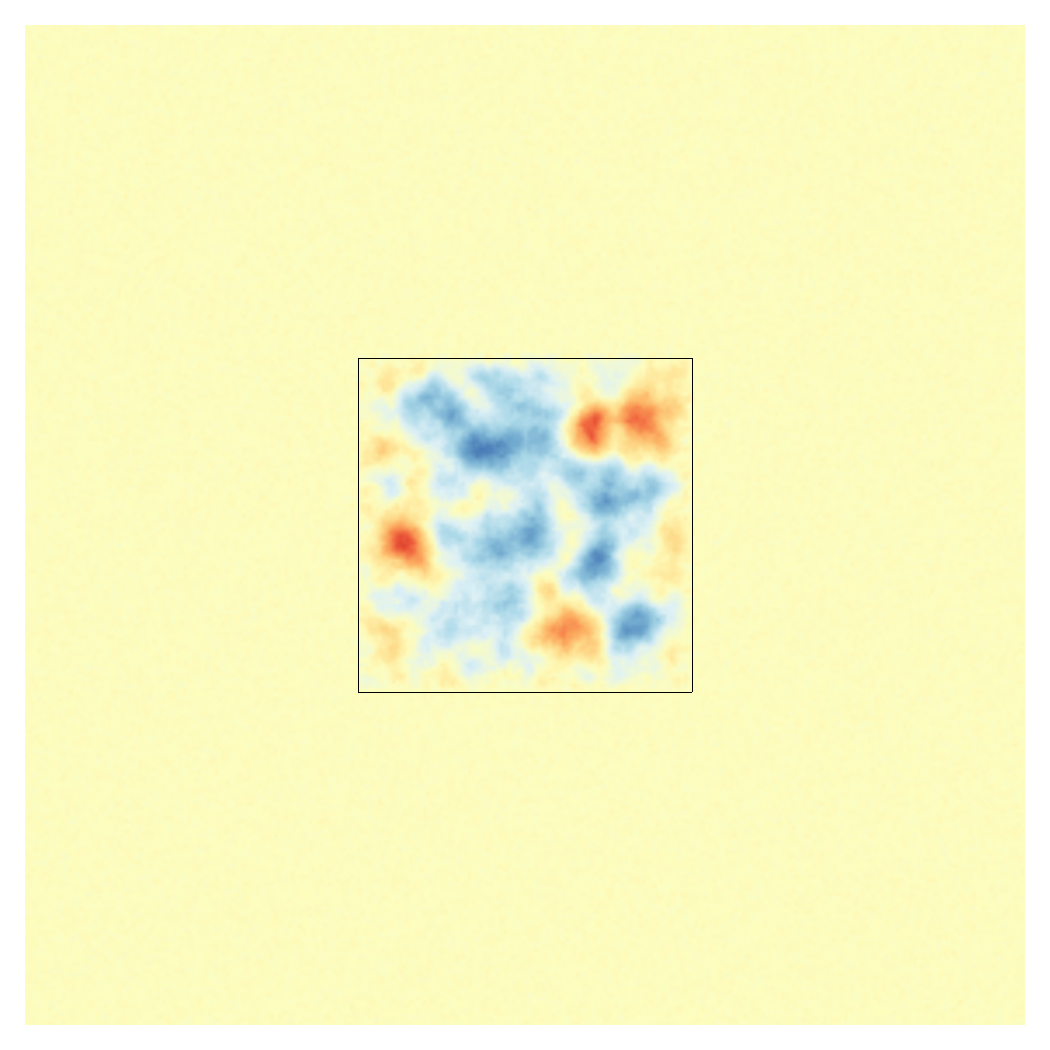}
\end{overpic}
\vspace{1em}
\begin{overpic}[width=0.32\textwidth]{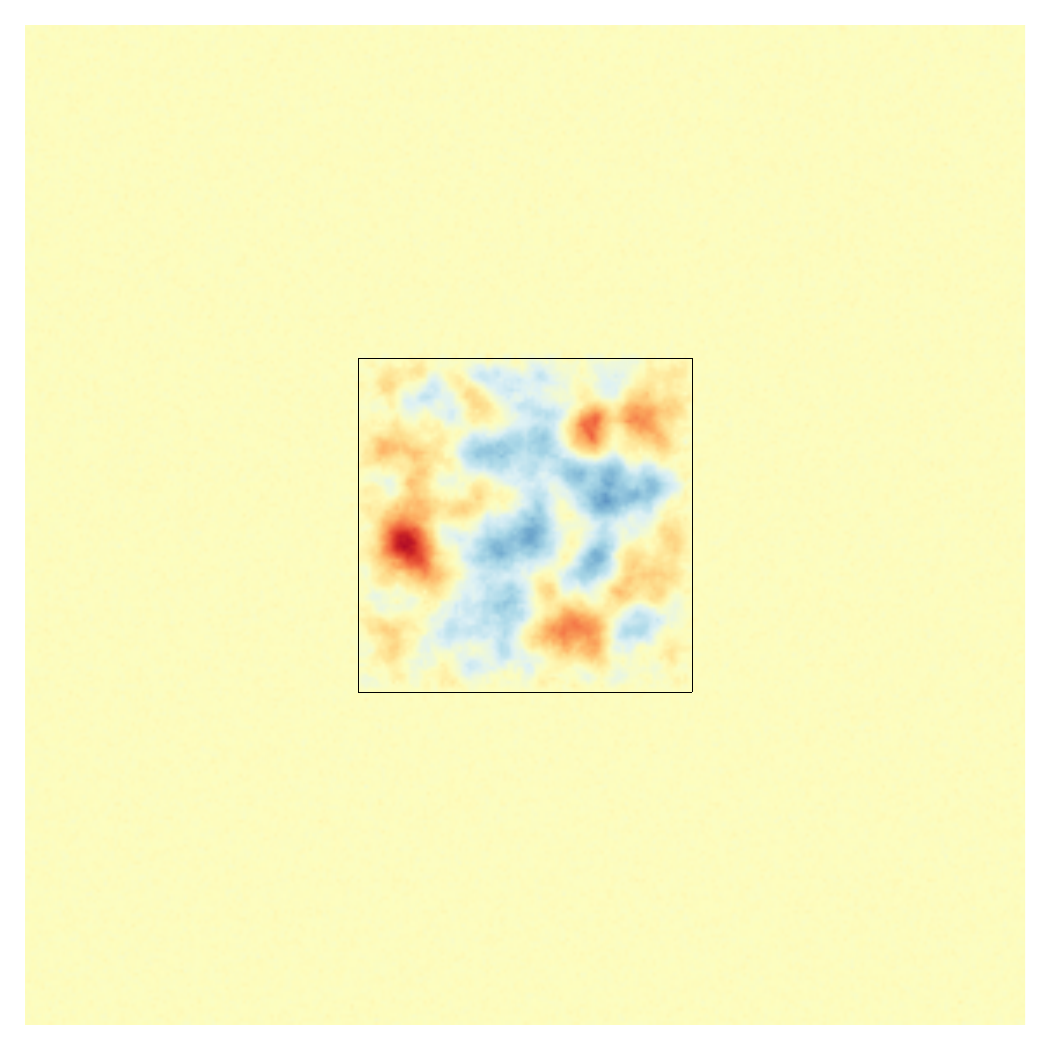}
\end{overpic}
\end{center}
\vspace{-0.6cm}

\caption{\label{fig:cmb_iters}
\emph{Top}: Reconstructed density field at given iteration. \emph{Bottom}: Difference of true density field with reconstruction at each iteration. Note that we have used the same mask/noise properties, as well as color scale, as in Sec \ref{subsec:CMBT}.}
\vspace*{-0.4cm}
\begin{center}
\includegraphics[width = 0.69\textwidth]{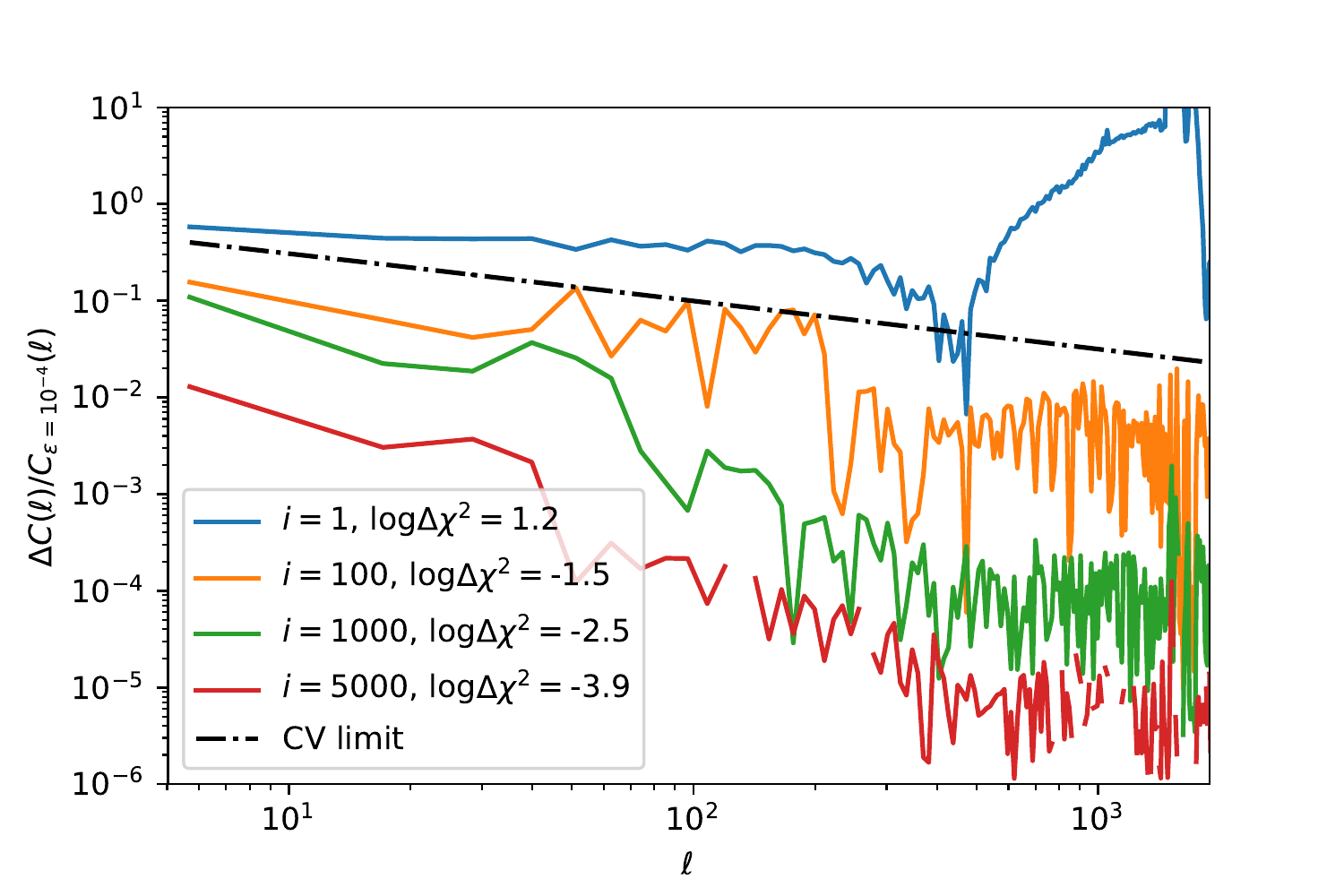}
\end{center}
\vspace*{-0.5cm}

\caption{\label{fig:cmb_scale_iter}
Change in convergence properties as a function of scale. We compare against the high convergence solution ($\epsilon = 10^{-4}$) rather than the true solution as the presence of noise will bias the end power spectra and calculating the noise bias and the Fisher information matrix (Hessian) for each step of the iteration would be computationally expensive.
}
\end{figure}

An important question to answer is to what the required convergence criteria are for a given algorithm/observable. In general, this will depend on what sort of scales are being probed and what other sources of error exist in the problem. In this section we will consider how changing the convergence criteria, $\epsilon \equiv \Delta \chi^2$, affects the net reconstructed map. We will specialize our analysis to that of the CMB case since it has the largest 
condition number, and most power on large scales, which will be particularly sensitive to reconstruction within the masked region. A similar analysis with our cosmic shear example will lead to smaller effects. 

To demonstrate the convergence properties of our technique we performed a high-accuracy run demanding $\epsilon < 10^{-4}$ as our convergence criteria, as opposed to $\epsilon < 10^{-1}$ for the runs in the main body of the paper. We show these results, as well as the difference with the true field, in Figure \ref{fig:cmb_iters}. Note that very quickly we find the true solution in the unmasked region, but continue to reconstruct the large scale modes in the masked region as the optimization rerouting continues.

We compare our convergence accuracy as a function of scale to the cosmic variance limit in Figure \ref{fig:cmb_scale_iter}. While in practice one wants the error on the reconstruction to be well below this limit, it provides a useful guideline for the necessary accuracy for reconstruction. Note that the properties of this reconstruction are a function primarily of the survey geometry; a  hypothetical full sky survey with no masked region and similar noise properties would converge much quicker to the optimal solution. Similarly, a case with smaller masked regions (for example only stellar masking) would find much faster convergence of the large scale modes.




\section{Joint E \& B Cosmic Shear Reconstruction}
\label{app:EB}

In the main text we only explored reconstructing the primary (i.e. curl-free) $E$-mode lensing potential of cosmic shear. However, there are various potential sources of $B$-mode effects within realistic observed lensing maps, such as instrumental effects, clustering of source galaxies \cite{bmodecluster}, and intrinsic alignments of galaxy shapes \cite{HirataSeljak04,Troxel14}. To control for these effects it is useful to perform a joint optimization of both $E$ and $B$ modes from the shear maps. The same tools could also be applied directly to the CMB polarization field from the $Q$ and $U$ maps \cite{HirataSeljak03recon}.

\begin{figure}[t]

\begin{center}
\begin{overpic}[width=0.495\textwidth]{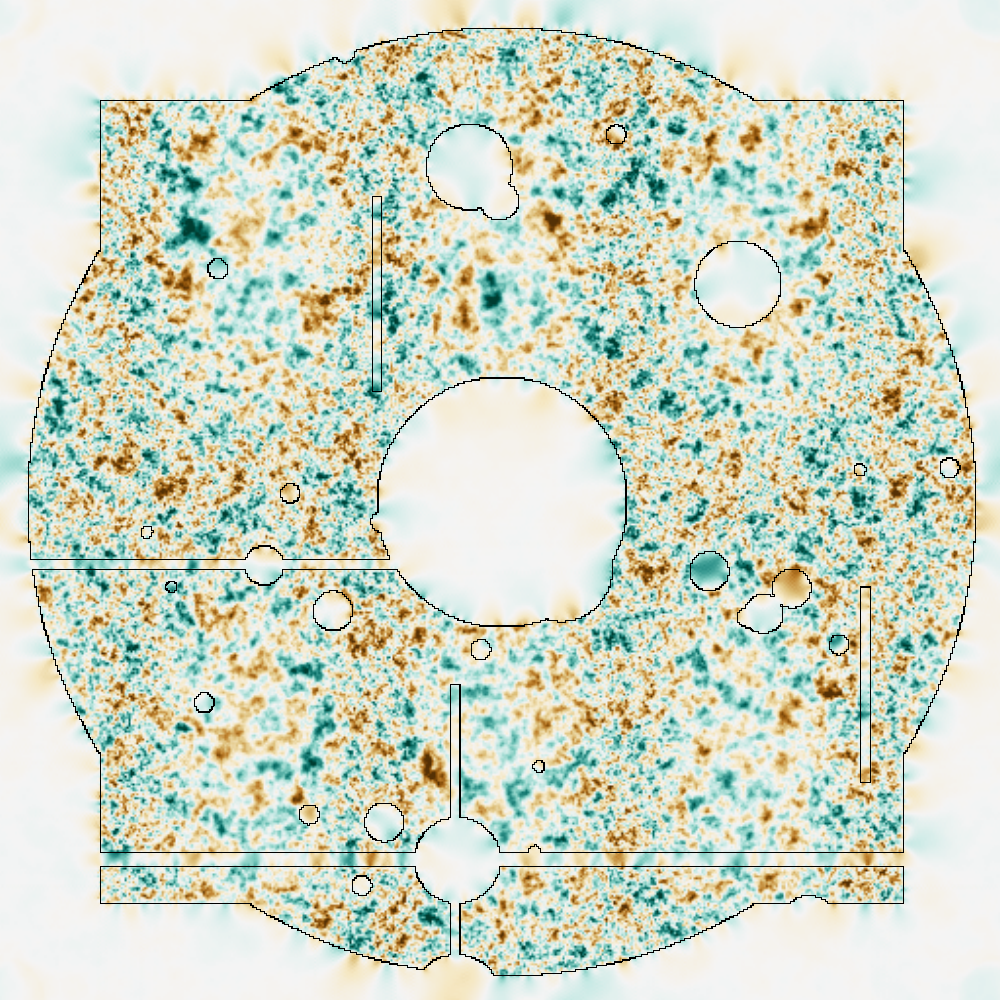}
\put(30,-10){\textsf{\scriptsize (a) Reconstructed E Field }}
\end{overpic}
\vspace{1em}
\begin{overpic}[width=0.495\textwidth]{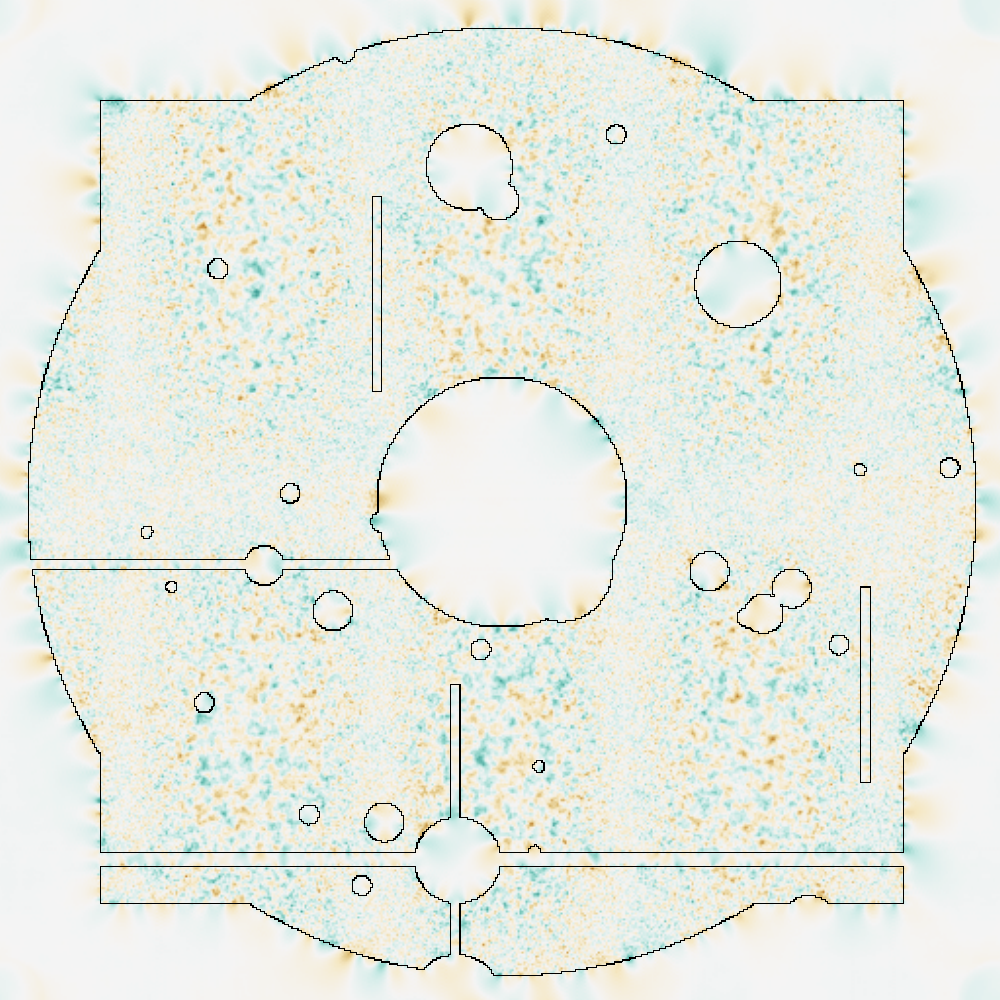}
\put(20,-10){\textsf{\scriptsize (b) Reconstructed B Field }}
\end{overpic}
\end{center}

\caption{\label{fig:EB_field}
Maximum likelihood E and B potential fields for cosmic shear case. Noise properties, mask, and color scale are the same as in Section \ref{subsec:cs}.
}
\end{figure}

The observed shear fields $\gamma_1$ and $\gamma_2$ can be expressed in terms of the E and B potentials as
\begin{eqnarray}
\left[\begin{array}{ c }\gamma_1  \\ \gamma_2  \end{array}\right] = \left[\begin{array}{ c c } (\partial_x^2 - \partial_y^2) & -2\partial_x\partial_y \\ 2\partial_x\partial_y & (\partial_x^2 - \partial_y^2) \end{array}\right] \left[\begin{array}{ c }\phi_E  \\ \phi_B  \end{array}\right],
\end{eqnarray}
where we assume flat sky. Our response matrix now takes two signal fields ($\phi_E$ and $\phi_B$) to two data fields ($\gamma_1$ and $\gamma_2$) and we perform the optimization over the signal fields. 

\begin{figure}
\begin{center}
\includegraphics[width = 0.85\textwidth]{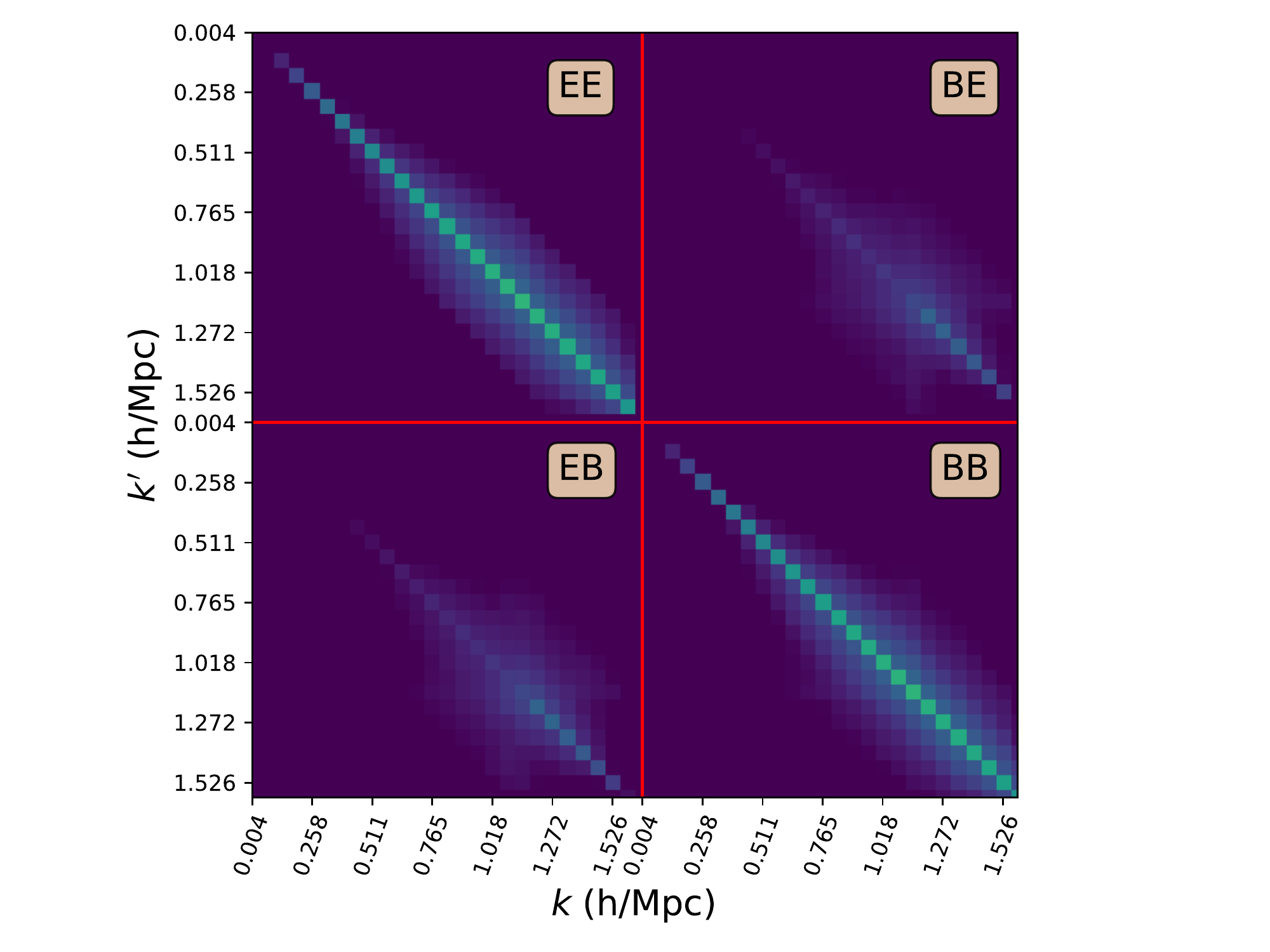}
\end{center}

\begin{center}
\includegraphics[width = 0.8\textwidth]{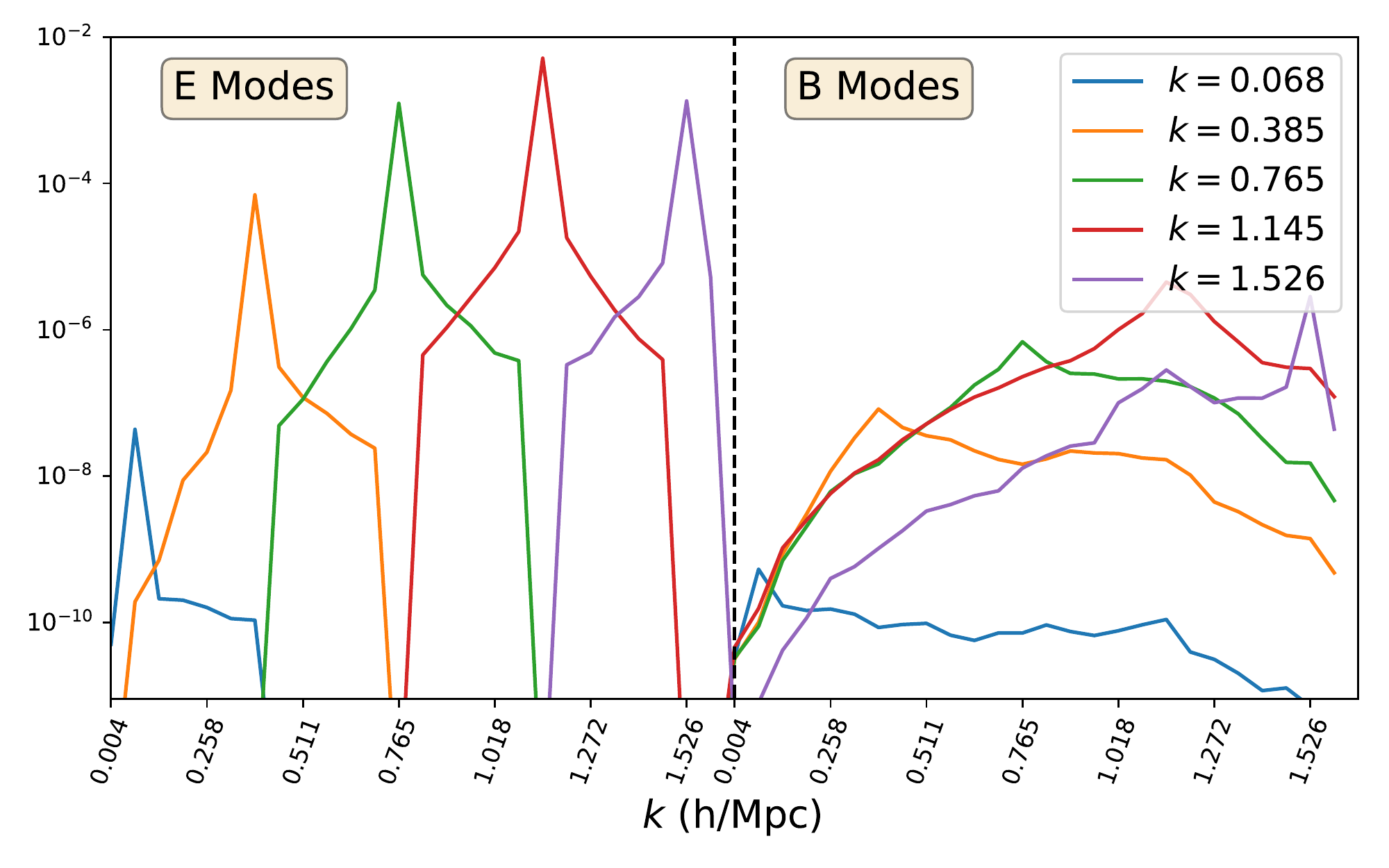}
\end{center}

\caption{\label{fig:EB_Fisher}
\emph{Top}: Full two dimensional Fisher matrix for the cosmic shear E/B joint reconstruction case. The matrix can be viewed as 4 blocks, with EE and BB the response of each type of mode to itself and the BE and EB reflecting the leakage between the modes induced by the survey geometry. \emph{Bottom}: Vertical cuts of the Fisher matrix. Note that to reduce numerical noise in the final reconstruction we have zeroed out terms in the EE blocks far from the diagonal. Since B modes are significantly sub-dominant in this example, a similar truncation of BB was not necessary.
}
\end{figure}

To study the joint reconstruction, we use the same starting E field as in Subsection \ref{subsec:cs}, but also induce a B field which has a power spectrum with the same shape as the E field but an amplitude $10^{-5}$ times smaller. We then do a joint reconstruction of both fields, yielding results shown in Fig \ref{fig:EB_field}.

\begin{figure}
\begin{center}
\includegraphics[width = 0.8\textwidth]{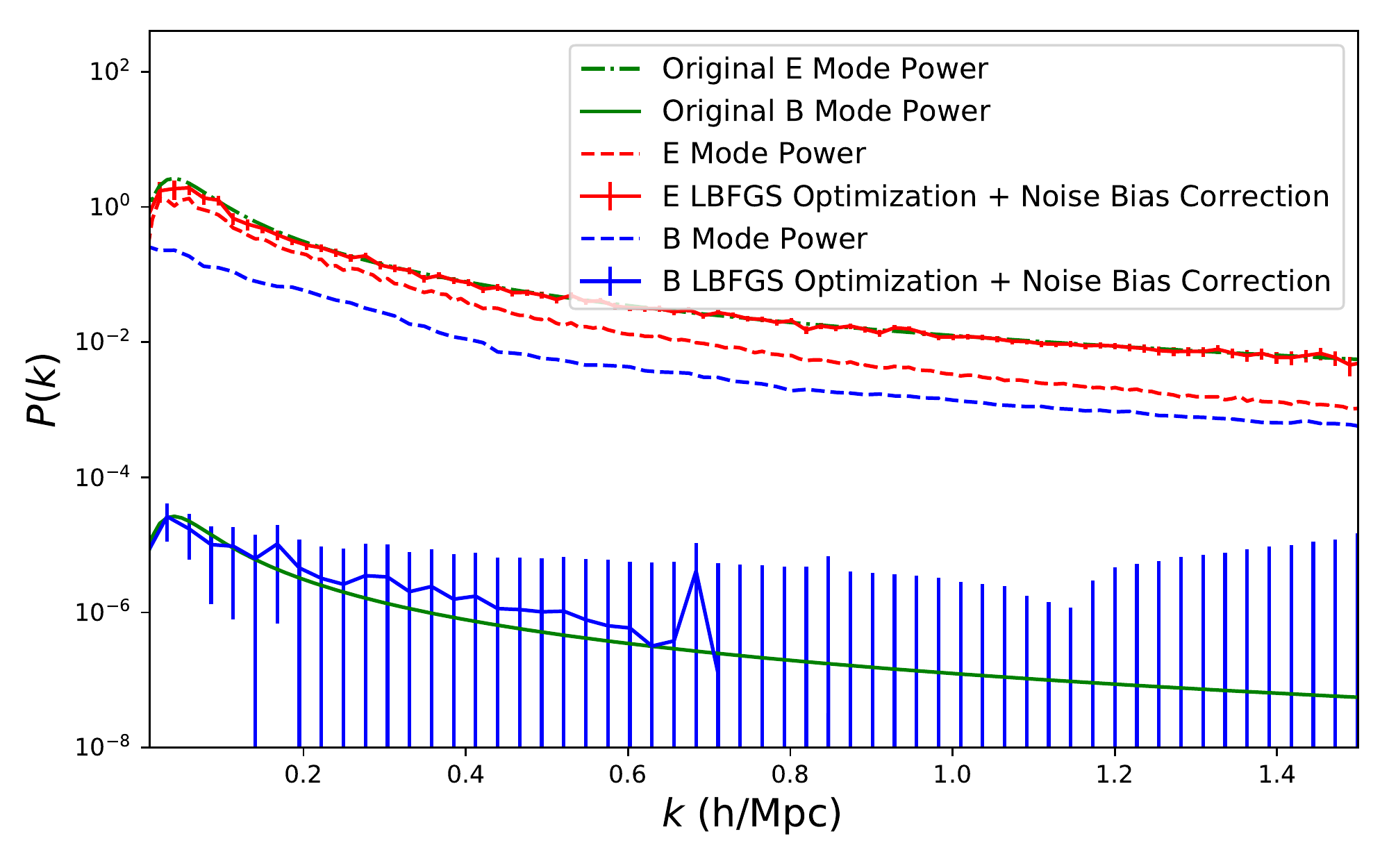}
\end{center}
\caption{\label{fig:EB_PS}
Reconstruction of the E and B power spectrum from mock observations using the maximum likelihood technique described in this work. Green lines indicate the original E/B power of the signal maps. The dashed red and blue lines indicate the power from the LBFGS optimized maps of E and B power respectively, while the solid lines indicate their MAP power spectra.
}
\end{figure}

The power spectrum estimation is now slightly more complicated as the Hessian matrix now has EE, BB, as well as EB, BE components to sum over, as shown in Fig \ref{fig:EB_Fisher}. The EB/BE components represent leakage between the channels, which in this case is dominated by E power leaking into B. We show this power spectrum reconstruction in Figure \ref{fig:EB_PS}. This is visually apparent in the reconstruction as we find an over-abundance of B power in the reconstructed map, which then gets down-weighted when this leakage is accounted for. In addition, as the B mode power is dominated by noise, it is difficult to accurately reconstruct its power from this one mock observation. 

Alternatively, rather than perform a full Fisher-like analysis for the B-mode power one could instead perform multiple realizations of the B-mode leakage and average them together to form a ``leakage bias" in analogous way as for the noise terms. This would have the possible advantage of requiring an additional optimization for each realization of the B-mode power, rather than an optimization for each mode of injected power as for the full Fisher analysis. If one is interested in studying many modes at once, treating the B-modes like noise bias would be computationally expedient.

\bibliographystyle{JHEP}
\bibliography{ms,cosmo,cosmo_preprints}
\end{document}